\documentclass{aa}  
\usepackage{graphicx}
\usepackage{natbib}
\newcommand{\teff}{\mbox{$T_{\rm eff}$}}
\newcommand{\logg}{\mbox{$\log g$}}

\newcommand{\degree}{\ensuremath{^\circ}}

\def\ms{\hbox{\,m\,s$^{-1}$}}         %m.s -1
\def\m2s2{\hbox{\,m$^{2}$\,s$^{-2}$}} %m2.s -2
\def\Msun{\hbox{$M_{\odot}$}}             %Msun
\def\Rsun{\hbox{$R_{\odot}$}}

\def \1s{$1\,\sigma$}

\def \t0{T$_0$}

\def\Rsun{\hbox{$R_{\odot}$}}

\newcommand{\mearth}{{\hbox{$M_{\oplus}$}}}
\newcommand{\rearth}{{\hbox{$R_{\oplus}$}}}

\usepackage{txfonts}
\begin{document} 

   \title{Revisiting the transits of CoRoT-7b at a lower activity level}

%   \subtitle{I. Overviewing the $\kappa$-mechanism}

   \author{Barros, S. C.  C. \inst{1},   Almenara, J.M. \inst{1}, Deleuil, M.   \inst{1}, Diaz, R.F. \inst{1,2}, Csizmadia, Sz.\inst{3}, Cabrera, J.\inst{3}, Chaintreuil, S. \inst{\ref{iparis}}, Collier Cameron, A. \inst{5},  Hatzes,A.\inst{6} ,Haywood, R. \inst{5}, Lanza, A. F. \inst{7}, Aigrain, S. \inst{8}, Alonso,~R.\inst{\ref{iiac},\ref{iull}}, Bord\'e, P.\inst{\ref{iias}}, Bouchy, F. \inst{1},  Deeg, H.J. \inst{\ref{iiac},\ref{iull}},  Erikson, A.\inst{3},  Fridlund, M.\inst{3}, Grziwa, S. \inst{\ref{iriuu}}, Gandolfi, D.\inst{7}, Guillot, T. \inst{\ref{ioca}}, Guenther, E.\inst{6}, Leger, A. \inst{\ref{iias}}, Moutou, C.\inst{1,14}, Ollivier, M.\inst{\ref{iias}}, Pasternacki, T. \inst{3},  P\"atzold, M.\inst{\ref{iriuu}}, Rauer, H. \inst{3,\ref{TU}}, Rouan,~D.\inst{\ref{iparis}}, Santerne, A.\inst{\ref{caup}}, Schneider,~J.\inst{\ref{iluth}} \and Wuchterl,~G.\inst{6}}

\authorrunning{S.C.C. Barros et al.}
   \institute{
 Aix Marseille Universit\'e, CNRS, LAM (Laboratoire d'Astrophysique de Marseille) UMR 7326, 13388, Marseille, France
              \email{susana.barros@lam.fr}
\and Observatoire Astronomique de l'Universite de Gen\`eve, 51 chemin des Maillettes, 1290 Versoix, Switzerland
\and Institute of Planetary Research, German Aerospace Center, Rutherfordstrasse 2, 12489 Berlin, Germany 
\and LESIA, UMR 8109 CNRS , Observatoire de Paris, UVSQ, Universit\'e Paris-Diderot, 5 place J. Janssen, 92195 Meudon cedex, France\label{iparis}
\and SUPA, School of Physics and Astronomy, University of St Andrews, St Andrews KY16 9SS, UK
%%%5 up 
\and Thuringer Landessternwarte, D-07778 Tautenburg, Germany 
\and INAF-Osservatorio Astrofisico di Catania, via S. Sofia, 78 - 95123 Catania. Italy
\and Department of Physics, University of Oxford, Denys Wilkinson Building, Keble Road, Oxford OX1 3RH 
\and Institut d'Astrophysique Spatiale, Universit\'e Paris-Sud \& CNRS, 91405 Orsay, France \label{iias}%
\and Instituto de Astrof\'isica de Canarias (IAC), E-38200 La Laguna, Tenerife, Spain\label{iiac}
%%10 up
\and Dept. Astrof\'isica, Universidad de La Laguna (ULL), E-38206 La Laguna, Tenerife, Spain\label{iull}
\and Rheinisches Institut f\"ur Umweltforschung an der Universit\"at zu K\"oln, Aachener Strasse 209, 50931, Germany\label{iriuu} 
\and Observatoire de la C\^ote d'Azur, Laboratoire Cassiop\'ee, BP 4229, 06304 Nice Cedex 4, France\label{ioca}
\and CNRS, Canada-France-Hawaii Telescope Corporation, 65-1238 Mamalahoa Hwy., Kamuela, HI 96743, USA 
%%11 up
\and Center for Astronomy and Astrophysics, TU Berlin, Hardenbergstr. 36, 10623 Berlin, Germany \label{TU}
\and Centro de Astrof\'isica, Universidade do Porto, Rua das Estrelas, 4150-762 Porto, Portugal \label{caup}
\and LUTH, Observatoire de Paris, CNRS, Universit\'e Paris Diderot; 5 place Jules Janssen, 92195 Meudon, France\label{iluth}
}

   \date{Received September ??, 2012; accepted March ??, ??}

% \abstract{}{}{}{}{} 
% 5 {} token are mandatory
 % contamination author name
  \abstract
{CoRoT-7b, the first super-Earth with measured radius discovered, has opened the new field of rocky exoplanets characterisation.
To better understand this interesting system, new observations were taken with the CoRoT satellite. 
During this run 90 new transits were obtained in the imagette mode. These were analysed together with the previous 151 transits obtained in the discovery run and HARPS radial velocity observations to derive accurate system parameters. 
A difference is found in the posterior probability distribution of the transit parameters between the previous CoRoT run (LRa01) and the new run (LRa06). We propose this is due to an extra noise component in the previous CoRoT run suspected to be transit spot occultation events. These lead to the mean transit shape becoming V-shaped. We show that the extra noise component is dominant at low stellar flux levels and reject these transits in the final analysis. We obtained a planetary radius,  $R_p=  1.585\pm0.064\,$ \rearth, in agreement with previous estimates.  Combining the planetary radius with the new mass estimates results in a planetary density of $  1.19 \pm 0.27\,  \rho_{\oplus}$ which is consistent with a rocky composition. The CoRoT-7 system remains an excellent test bed for the effects of activity in the derivation of planetary parameters in the shallow transit regime.}

\keywords{planetary systems -- stars: individual: (CoRoT-7b) --stars:activity --techniques: photometric--methods:data analysis--methods:observational }

\maketitle
%
%________________________________________________________________

\section{Introduction}
CoRoT-7b was the first transiting super-Earth discovered \citep{Leger2009} and it is one of the most interesting planets detected by the Convection Rotation and planetary Transits (CoRoT) space telescope. CoRoT-7b was detected in the LRa01 run of CoRoT from the 24 October 2007 to the 3 March 2008.
The planet orbits a G9V type star every $\sim 0.85\,$days \citep{Leger2009}.
The host star was found to be young (1.2-2.3 Gyr) and active showing a 2\% amplitude variability in the CoRoT light curve. The stellar activity severely affected the radial velocity (RV) follow-up observations needed to confirm the planet and to derive its mass \citep{Queloz2009}. During the RV observations the semi-amplitude of activity-induced RV variability was  20 m/s and hence, higher than the planet signature $\sim 1.6-5.7\,$m/s. Several methods were developed to correct the activity-induced RV variability. However, the measured RV planet signature was found to be method dependent with estimates of the mass of CoRoT-7b ranging from $2.26$ to $8.0\,$\mearth\ \citep{Queloz2009,Hatzes2010, Hatzes2011,Boisse2011,Pont2011, Ferraz-Mello2011}. Nevertheless, 90\% are consistent with a  mass $6.2 \pm 1.2 $ \mearth\ .
 Interestingly, two additional planetary signatures were seen in the RVs one with a period of $3.69$ days and a mass of $8.4\pm 0.9 \mearth$ \citep{Queloz2009} and a second companion with period $9.0$ days and mass of $16.7\,$\mearth\ \citep{Hatzes2010}.

It was suspected that the stellar activity also affected the derived transit parameters of CoRoT-7b \citep{Leger2009}. The stellar density derived from transit fitting was found to be much lower ($\rho_*= 0.12 \rho_{\odot}$) than the derived spectroscopic value ($ \rho_* = 1.4 \rho_{\odot}$). However, the authors noted that when subsets of 4-5 transits were fitted separately, the resulting stellar density was in agreement with the expected value for a G9V type star. Therefore, they concluded that adding all the transits together degraded the transit shape and suggested two possible causes. The first was transit timing variations due to hypothetical additional planets in the system and the second was stellar variability. To avoid the underestimation of the stellar density, the spectroscopic derived radius was used as a prior in their transit fitting. Later, an in-depth spectroscopic study of the host star improved the stellar radius and consequently the planetary radius which was found to be $1.58 \pm 0.10\,$ \rearth\   \citep{Bruntt2010b}.

In this paper we test the effect of stellar activity in the transit observation of CoRoT-7b. Stellar activity can affect the transit light curves in two different ways \citep{Czesla2009}. Dark spots or bright faculae outside the transit chord (projected path of the planet in the stellar surface) alter the out-of-transit stellar flux and can also introduce trends in the normalisation. Spots or faculae inside the transit chord (planet-spot occultations) will affect the transit shape. Both effects will influence the parameter estimation. 

The out-of-transit stellar variability of CoRoT-7b during the LRa01 CoRoT observations was modelled by \citet{Lanza2010} who found that the photometric variability is dominated by cool spots with reduced facular contribution in comparison to the sun. This was also confirmed by ground-based colour-photometry acquired in Dec 2008 to Feb 2009 \citep{Queloz2009}. Moreover,  \citet{Lanza2010} found that in CoRoT-7 there were three main active longitudes with rotation periods ranging from 23.6 to 27.6 days, which were attributed to differential stellar rotation. These active longitudes could persist for several years as suggested  by the phase coherence of the light modulation between LRa01 CoRoT light curve and the light curve of \citet{Queloz2009}. The mean lifetime of individual active regions is about 18 days and they cover a maximum of $1.6\%$ area of the stellar surface. Furthermore, an analysis of the spot model residuals lead \citet{Lanza2010} to propose that there is a population of small active regions with lifetime of 4-5 days that cover $\sim0.1$ percent of the stellar surface.% These can result in a overestimation of the planetary radius.

Planet-spot occultations were first observed in the transit of HD209459b by \citet{Deeg2001,Silva2003}. For example in CoRoT-2b \citep{Alonso2008}, the analysis of the evolution of the spots through several transits allowed probing the stellar surface and deriving spot properties \citep{Lanza2009, Wolter2009,Silva-Valio2010,Silva-Valio2011}. However, in ground based observations where only a few transits are observed, the effects of spots can be difficult to recognise and to correct \citep{Barros2013}.

To better understand and characterise this iconic system CoRoT-7 was re-observed with the CoRoT satellite for additional $80\,$days, from the 10th of January 2012 to the 29th of March 2012.  Furthermore,  simultaneous RV observations were taken with HARPS in the first 26 consecutive nights of the CoRoT observation.
In this paper we present the results of the analysis of the new CoRoT photometric observations. These new data allowed a better understanding of the effect of stellar activity in shallow transits and updating the system parameters. Although the RV analysis will be presented elsewhere (\citealt{Haywood2014}, Hatzes, A. et al. in prep), the full HARPS RV set is included  in our analysis in order to obtain consistent parameters. We begin by presenting the data reduction procedure in Section 2 where we also describe the CoRoT imagette pipeline. In Section 3 we present our transit fitting model and show our results for the new LRa06 data in Section 4. In Section 5 we compare the two CoRoT observations, LRa01 and LRa06. The tests performed to LRa01 to investigate the distortion of the parameter posterior distributions that lead to rejecting some transits are explained in Section~6. The final results presented in Section~7 are discussed in Section~8. Finally the conclusions are given in Section~9.

%__________________________________________________________________
\section{Data reduction}
\label{data}
The CoRoT satellite has two science channels for its two science goals: asteroseismology and exoplanet search.
 In the exoplanet channel $6\,000-12\,000$ target stars are monitored in each run and due to limitations on telemetry, data reduction is performed on board and only the light curves are transmitted to the ground. However, for 40 targets per channel, a window of $15 \times10 $ pixels is transmitted to the ground. These are called imagettes and are used for special targets or for the bright stars in the field. While there are two sampling rates in the light curve mode: 512 seconds and 32 seconds, in the imagette mode the sampling rate is always 32 seconds.
The imagette mode has several advantages compared to the light curve mode and hence, the CoRoT-7 re-observation was performed in the imagette mode. In LRa01 CoRoT-7 was observed in the light curve mode with a sampling rate of 32 seconds.
A full description of the CoRoT satellite can be found in \citet{Auvergne2009}.

\subsection{Imagette pipeline}

The imagette pipeline is composed of two parts. The first is related to image calibration and consists of three main corrections: the bias correction, the cross talk effect \footnote{Perturbation due to simultaneous reading of the stellar seismology channel \citep{Auvergne2009}.} correction,  and the background correction. To estimate the latter, there are 196 background windows spread over the CCD each consisting of $10 \times 10$ pixels. In the beginning of the mission, the background was evaluated to be very uniform over the CCD and therefore the median of the background windows was used for the correction. However, recently it was found that due to the ageing of the CCD the background is no longer uniform and depends on the position along the reading direction of the CCD (y-direction). This is mainly due to an increase in the dark current and a decrease of the transfer efficiency (Chaintreuil  et al. in prep.). Consequently, a correction which depends on the y-position of the target in the CCD was added to the median in the estimation of the background. This correction was included for both observations of CoRoT-7 presented here.

In the second part of the pipeline, the extraction of the light curve from the image time series is performed using aperture photometry. The procedure starts by computing the aperture that maximises the signal-to-noise in the mean image. Subsequently the centroid of each image is computed.
 The centroid position is used to centre the images before extracting the flux inside the optimum aperture. Since there is a prism in the exoplanet channel in front of the CCD that produces a low resolution spectrum, colour information can also be extracted. The imagette pipeline provides three colour light curves for the red, green and blue parts of the spectrum, however, these colours do not correspond to the standard photometric filters. The number of columns in the dispersion direction used for each colour is calculated so that the red flux is $40 \pm 12$ \%, the blue flux is $30 \pm 8$ \% of the total flux and there is at least one column for the green \citep{Rouan1999}.
An important  part of the pipeline is the flagging of the bad data points mainly due to passage through the south Atlantic anomaly (SAA), cosmic ray events and hot pixels.

Recently, the computation of the barycentre was also included in the imagette pipeline. 
The barycentre is computed by subtracting the mean position of the stars in the field (CoRoT line of sight) from the position of the centre of light of the target star. The barycentre is very sensitive to the existence of contaminant stars inside the aperture \citep{Barros2007} which can be used to discard false positives as it is done for the Kepler candidates (e.g. \citealt{Borucki2011}).

\subsection{Customised data reduction}
\label{custom}
\begin{figure}
  \centering
  \includegraphics[width=\columnwidth]{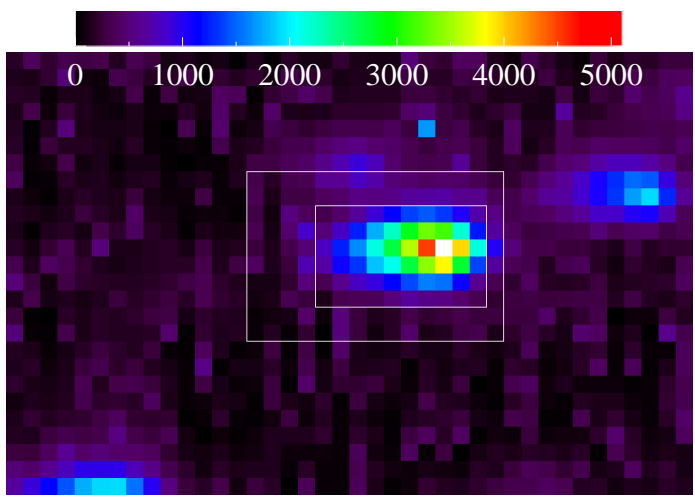}
  \caption{Field of the CoRoT CCD around CoRoT-7. The bigger rectangle shows the imagette position while the smaller rectangle is the optimum aperture used to obtain the final light curve. The images are shown in a square root scale of the flux so that the contaminant in the top left of CoRoT-7 becomes visible. }
  \label{imagette}
\end{figure}

During the CoRoT run LRa06 which started on the 10th of January 2012 and was finished on the 29th of March 2012, CoRoT-7 was re-observed in the imagette mode. The most obvious advantage of the imagette mode is that the data reduction can be optimised and customised for a particular target. In Figure~\ref{imagette}, we present an image of the field of view around CoRoT-7 where we show the position of the imagette (larger rectangle) on the CoRoT CCD.

To improve the data reduction, we started by optimising the target aperture in order to minimise the rms of the final light curve on time scales shorter than 30 minutes. Variability on time scales longer than 30 minutes is dominated by intrinsic stellar variability while variability on shorter timescales is dominated by instrumental and reduction noise. From our signal-to-noise optimisation procedure, we realised that due to jitter noise and the elongated PSF shape, rectangular shapes were preferred and pixels in the border of the imagette should be avoided. The optimum aperture found is the smaller rectangle of $10 \times 6$ pixels shown in Figure~\ref{imagette}. The contaminant seen on the top border of the imagette was deliberately excluded from the aperture, resulting in a zero estimated contamination inside the final aperture. The final step of the data reduction was the removal of the light curve outliers that deviate from the local mean by more than $5 \sigma$. 

\begin{figure}[htp]
  \centering
\begin{tabular}{c}
\includegraphics[width=\columnwidth]{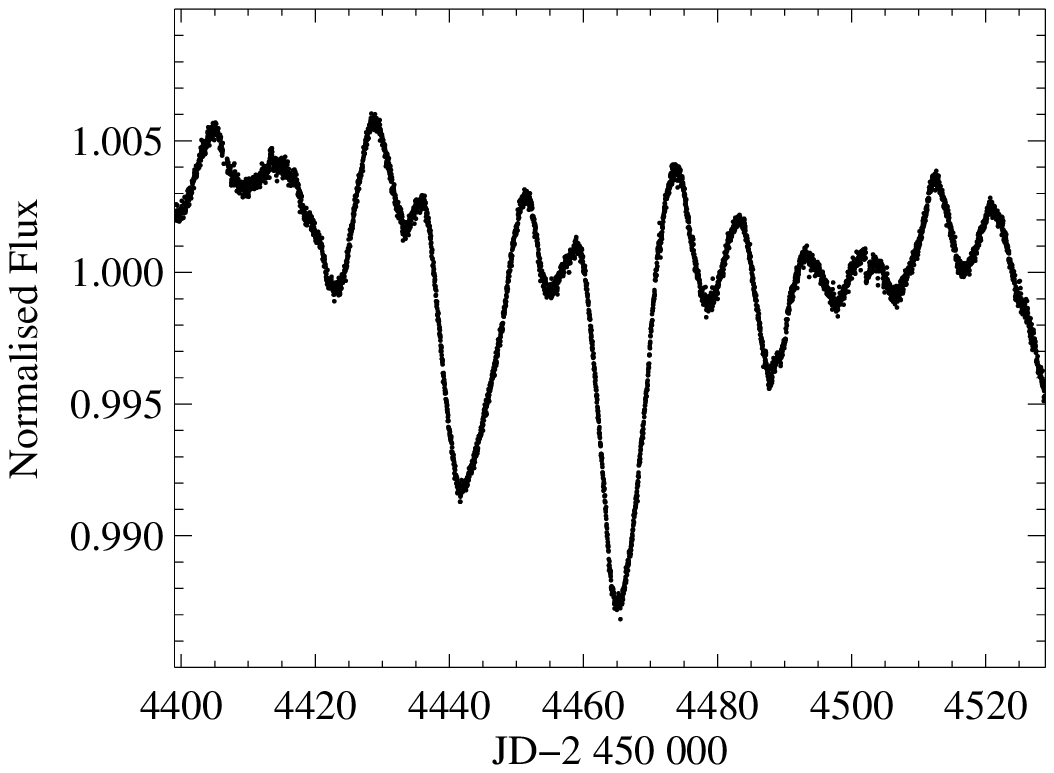}\\
\includegraphics[width=\columnwidth]{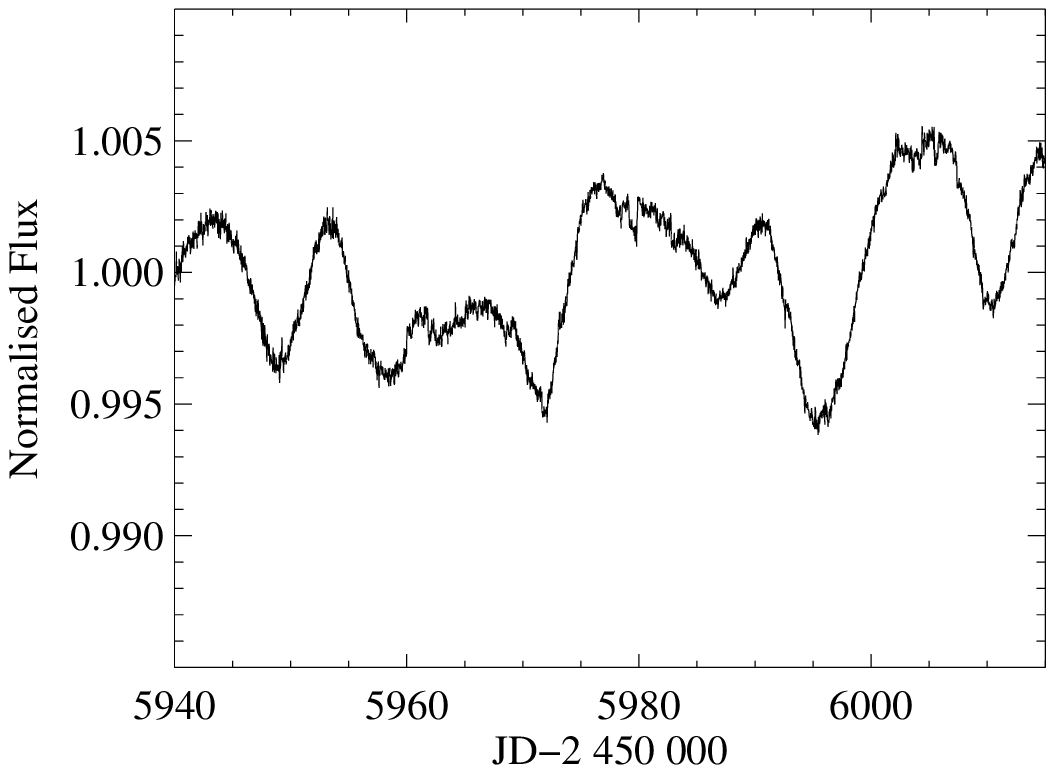}\\
\end{tabular}
\caption{Light curve of CoRoT-7b binned to 30 minutes for clarity during LRa01 (top) and LRa06 (bottom).  It is evident that the variability level during LRa01 is twice of the one in LRa06 CoRoT light curve.}
\label{lc}
\end{figure}

The customisation of the data reduction led to a 50\% decrease of the light curve's rms relative to the automatic pipeline. The final light curve after sigma clipping is shown Figure~\ref{lc}. It has a mean rms of 1300 ppm per point, i.e.  $\sim 1.3 \times$ the expected photon noise. The rms is higher than in the previous run probably due to the ageing of the CCD an increasing of the dark current and a decreasing of the charge transfer efficiency. To reach the photon noise level, as in this case, it is very important to reduce sources of systematic effects for example due to movement of the target across the CCD \citep{Barros2011b}.
The guiding of the CoRoT satellite is based on bright stars in the seismology field and is very accurate, resulting in a very stable stellar centroid. For example, the centroid of CoRoT-7 varied by 0.2 pixels in the x-direction (dispersion direction) and 0.09 pixels in the y-direction during the 80~days of the run.

We included in our transit analysis procedure the previous transits of CoRoT-7b obtained in the LRa01 run and presented by \citet{Leger2009}. We used the latest version of the light curve pipeline to re-reduce the previous data that includes the background correction in the y-direction mentioned above. The aperture used for the LRa01 data is larger than the one for LRa06. It has a total of 96 pixels and a shape closer to the CoRoT-7 point spread function. This results in a contamination value of $0.92 \pm 0.57$ \%\  \citep{Gardes} which is slightly higher than for LRa06 but not significantly so. 

For the transit analysis we extracted individual transit light curves with a length of $\sim 210$ minutes which corresponds to approximately three transit durations, centered at the mid-transit time. The original sampling of the light curves (32 seconds) was kept and no binning was applied. Each transit was normalised by a local baseline function. We tested a linear versus quadratic baseline function and found that a quadratic baseline function was not necessary. In total 151 transits of CoRoT-7b were analysed in LRa01 and 90 in LRa06.
For the final analysis the uncertainties of each of transit light curve were scaled to account for the red noise. We estimated the time corrected noise $\beta$ using the procedure of \citet{winn2008}. For LRa01 the mean of the beta of each transit is 1.56 and beta ranges from 1.1 to 2.23. For LRa06 the mean is 1.04, the maximum is 1.33 but most of the values are 1.

\section{Transit Model}
\label{modelfull}
The transits observed by CoRoT in the runs LRa01 and LRa06 were modelled simultaneously with the radial velocities from HARPS with the PASTIS code \citep{Diaz2014}. The HARPS radial velocities comprise the previous observations presented by \citet{Queloz2009} and recent observations described by \citet{Haywood2014}. As mentioned before, the recent RV observations are simultaneous with the LRa06 CoRoT observations. To obtain consistent system parameters, PASTIS uses a Markov chain Monte Carlo (MCMC) algorithm to sample the parameter's posteriors. The radial velocities are modelled by a keplerian orbit, the transit light curves are modelled with the EBOP code \citep{Etzel1981,Popper1981} and the stellar parameters are interpolated from Geneva \citep{Mowlavi2012}, Dartmouth \citep{Dotter2008} or PARSEC \citep{Bressan2012} stellar evolution tracks. At each step of the chain, the proposed stellar density is combined with the stellar metallicity, stellar temperature and the stellar tracks in order to derive the stellar mass and radius. Consistent quadratic limb darkening parameters are also derived from the tables of \citet{Claret2011} at each step. To minimise the effect of stellar variability and of other possible planets on the measured RVs we used a method similar to \citet{Hatzes2010, Hatzes2011}. We considered only nights with more than 1 RV observation per night and we included a RV offset for each night which was fitted simultaneously with the system parameters. This is similar to a low pass filter, removing variability on time scales longer than 1 day.

The fitted parameters in our full model are: the orbital period P, the transit epoch T0, the stellar reflex velocity K, the orbital eccentricity e, the longitude of the periastron $\omega$, the inclination $i$, the ratio of planet radius to star radius $R_p/R_*$, the stellar density $\rho_*$,  53 nightly radial velocity offsets, the out of transit flux for each CoRoT run and jitter noise for each CoRoT run. Uniform priors were used for all the parameters except for the relative planet to star radius, $R_p/R_*$, the inclination and stellar parameters. For $R_p/R_*$ the Jeffreys' prior was used while for the inclination a sine prior was used to impose an isotropic distribution of orbit orientations. Normal priors were used for the stellar density, metallicity and temperature according to the values  of  \citet{Bruntt2010b} ( \logg $=4.47\pm 0.05 $, [Fe/H] $= 0.12 \pm 0.06$,  \teff $ = 5250 \pm 60  \,$K ). The age of CoRoT-7 was constrained to be less than $3$~Gyr as in previous analysis \citep{Leger2009, Bruntt2010b}. 
For the final results 30 MCMC chains for each of the three stellar models were combined. Each MCMC is comprised of $400\,000$ steps and was started at random points drawn from the joint posterior. PASTIS uses a principal component analysis that improves mixing while convergence is tested with \citet{Gelman92} statistic \citep{Diaz2014}.

\subsection{Pure geometric model}
\label{modelgeo}
The analysis of the CoRoT-7 full data set proved to be complex because of the low signal-to-noise transit signal. The transit depth is $\sim 0.00034$ which is approximately one third of the rms of the light curve. Therefore, the transits need to be combined to derive the transit shape which implicitly assumes a constant transit shape and a linear ephemeris. To test these assumptions and gain insight into transit parameter derivation for low signal-to-noise transits in the presence of stellar activity we performed several tests.
For these tests, we included only the transit light curves in the PASTIS fit, which we refer to as a pure geometric fit. In this fit the fitted parameters are: three shape parameters ( $R_p/R_*$, the inclination ($i$) and the normalised separation of the planet ($a/R_*$)), the two ephemeris parameters (P and T0), the out-of-transit flux and the jitter.  For simplicity, we kept the limb darkening coefficients fixed to the values of \citet{Leger2009} and assumed a circular orbit. Hence, $a/R_*$ represents the normalised separation of the planet at the time of the transit and needs to be corrected in a first order approximation by $\frac{1+e\sin\omega}{\sqrt{1-e^2}}$, for eccentric orbits. In the next sections we describe the most enlightening of our tests before we present our results in Section~\ref{results}.
%Other tests were also preformed, specificly we attempted simple data filtering but found 

\section{Analysis of LRa06 new CoRoT data}

 We begin by performing a pure geometric fit to the LRa06 transit light curves. This results in $a/R_*=5.90^{+0.05}_{-2.8}$, $i= 90_{-10}$\degree\ and $R_p/R_*=0.01702^{0.0023}_{-0.00052}$ which implies a stellar density of $3.79^{+0.12}_{-3.2} \rho_{\odot}$. This is higher than the stellar density derived from spectroscopy but consistent within the errors. However, our transit derived stellar density contrasts with
 the unconstrained analysis of LRa01 \citep{Leger2009} that led to a transit derived stellar density lower than expected from spectroscopy as mentioned above.  To solve this inconsistency, in the analysis of LRa01 the stellar radius was used as a prior in the transit fit which resulted in the final solution $a/R_*= 4.27 \pm 0.20$, $i= 80.1 \pm 0.3\degree$, $R_p/R_*=0.0187 \pm 0.003$ \citep{Leger2009}. This is consistent with our results within $1 \sigma$.
 
We find that, for LRa06, the inclination is poorly constrained, with the $2\sigma$ interval ranging from 68\degree\ to 90\degree, i.e. all possible values that would produce a transit. This suggests that due to the low signal-to-noise of the transit, the transit shape of CoRoT-7b is not well resolved leading to a degenerate solution. This hypothesis is tested in the next subsection.

\subsection{White noise test}
\label{whitenoise}
To test the posterior distribution of transit parameters derived from low signal-to-noise light curves we simulated transits using a preliminary full PASTIS fit (presented in Section~\ref{results}).
The transits were simulated using the \citet{Mandel2002} transit model parametrised by $a/R_*=4.08$, i$=79.20$\degree, $R_p/R_*=0.01781$, the limb darkening coefficients $\gamma_1=0.4396$ and $\gamma_2=0.2598$ and assuming a linear ephemeris (period of $0.85359199\,$days and epoch of $2454398.07669$). This transit model shape will be used in all the tests presented here.

For observed times within the same transit windows as the data analysed above, we simulated transits and added white Gaussian noise with standard deviation equal to the standard deviation of the real data, $\sigma=0.0013$.

\begin{figure}
  \centering
  \includegraphics[width=\columnwidth]{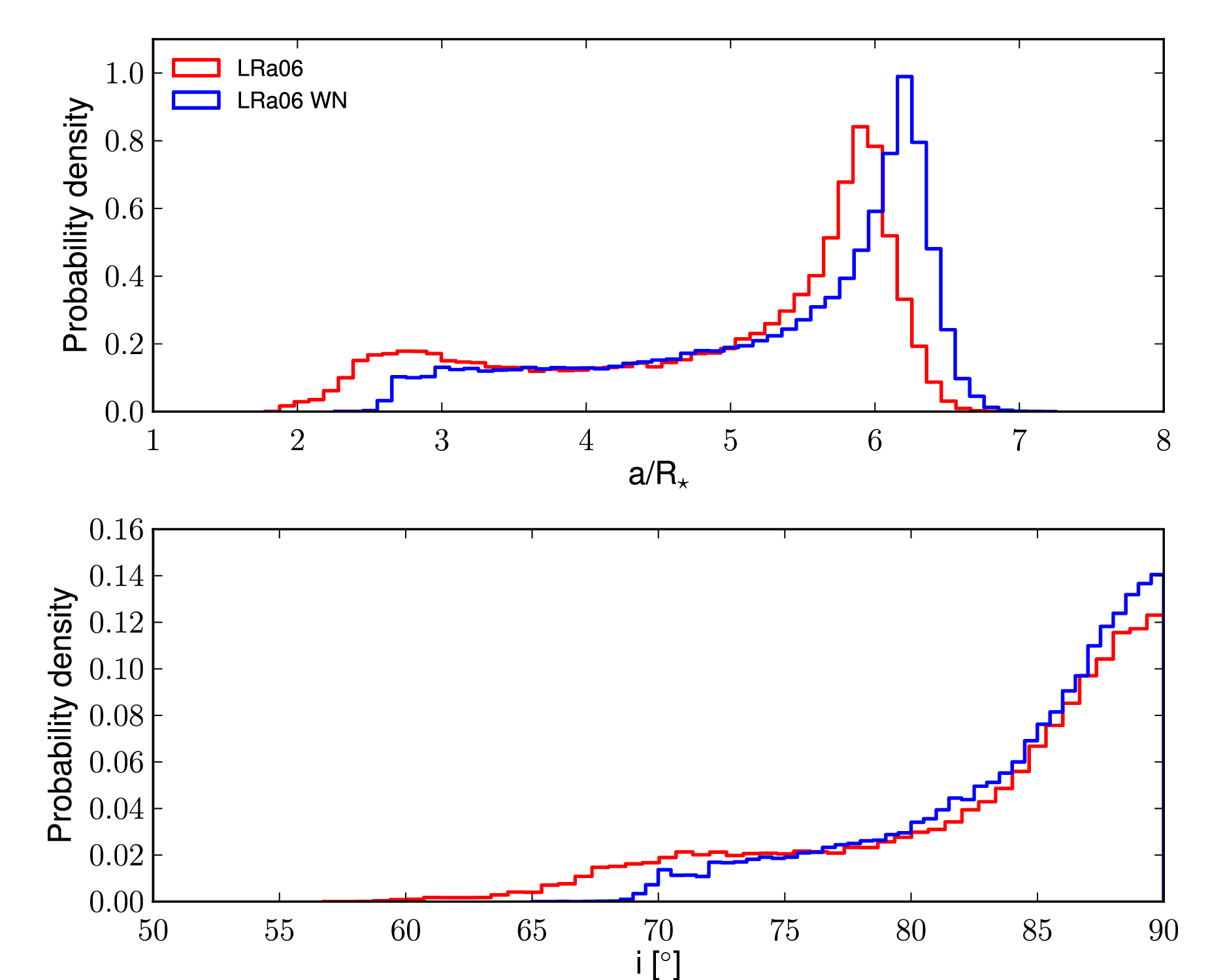}
  \caption{Posterior distribution of the parameters, $a/R_*$ and $i$ for the PASTIS pure geometric fits of LRa06 (red) and the transit model plus white noise simulations at the observing times of LRa06 (blue).}
  \label{lra06posteriors} 
\end{figure}

We performed a pure geometric fit to the white noise simulated transits following the same procedure as for the real data.
The posterior distributions of the problematic fitted parameters, $a/R_*$ and $i$ are shown in Figure~\ref{lra06posteriors}. We find that the posterior distribution of the real and simulated transits are similar which confirms our hypothesis that the transit shape is not well constrained particularly the ingress/egress time which results in an unconstrained inclination.
Therefore, the LRa06 posterior distribution is characteristic of small planets with poor signal-to-noise of the light curve. Within $2 \sigma$ we can only constrain the inclination to be higher than 68\degree\  i.e. all the values that would produce a transit are allowed. We note that the specific shape of the distribution might depend on the specific parametrisation of the transit shape and priors.
In conclusion, the LRa06 parameter posterior distributions are consistent with the final results presented by \citet{Leger2009}.
Moreover, for low signal-to-noise transits the transit-derived stellar density is not well constrained and the stellar density derived from spectroscopy or asteroseismology can help constrain the transit geometry as we explore next.

\subsection{Stellar density}

In this section we discuss in detail our treatment of the stellar density which constrains the planetary parameters. In particular, we explain the differences between our transit fitting method implemented in PASTIS and the method applied by \citet{Leger2009}. 

Usually, for high signal-to-noise transit light curves the stellar density is derived directly from the transit \citep{Seager2003} and at better precision than the \logg\ estimated from spectroscopy. Therefore, the stellar density derived from the transit together with \teff\ and [Fe/H] estimated from spectroscopy can be combined with the stellar models \citep{Sozzetti2007} or empirical calibrations \citep{Torres2010} and used to constrain the stellar mass and radius (e.g. \citealt{Barros2011b}). Interesting tests on using the transit derived density to better constrain stellar parameters can be found in \citet{Torres2012,Yilen2013}.  However, for low signal-to-noise transits where the constraint on the stellar density by the transit is weak, it can be advantageous  to use the spectroscopic derived \logg\ to estimate the stellar mass and radius, which consequently constrains the transit parameters. This was the method applied by \citet{Leger2009} for CoRoT-7b, who included the spectroscopic derived radius as a prior in the transit analysis. This was necessary to solve the density inconsistency that was mentioned before.
Within PASTIS the transits are modelled together with the stellar models and the stellar properties (stellar density, [Fe/H] and  \teff), which is equivalent to a prior in the stellar properties but works in a self-consistent way, appropriately weighting the constrains on the density.
 
\begin{figure}
  \centering
  \includegraphics[width=\columnwidth]{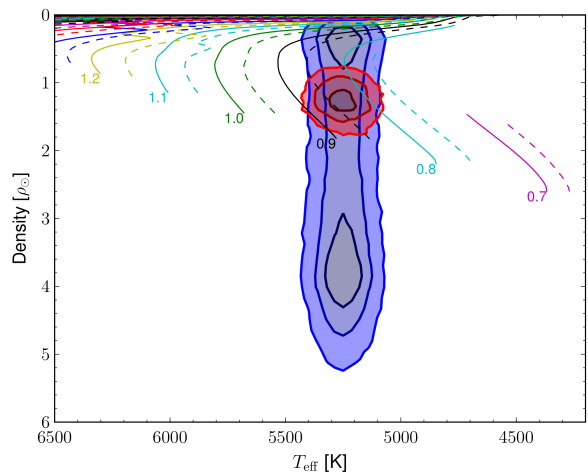}
  \caption{Dartmouth \citep{Dotter2008} stellar evolution tracks for stellar metallicities of 0.05 (solid lines) and 0.15 (dashed lines). The posterior distribution  $1\sigma$, $2\sigma$ and $3\sigma$ contours for a pure geometric fit of LRa06 is shown in blue. Including the stellar models in the fit allows to better constrain the density by reducing the posterior distribution to the red area. }
  \label{density}
\end{figure}

In Figure~\ref{density}  we show the density derived from the fit of LRa06 transit data including the stellar models and excluding the stellar models. We also show the stellar tracks for stars with masses from 0.6 to $1.3$\Msun\ with [Fe/H]$=0.05$ and [Fe/H]$=0.15$. It is evident that the stellar density derived from the transit is poorly constrained with the $2\sigma$ interval including $0.2-4.7  \rho_{\odot}$. This is due to the short ingress/egress time for small-sized planets combined with the low signal-to-noise of the light curve. In this case, stellar evolution tracks together with \logg,  \teff\ and [Fe/H] estimated from spectroscopy add information to our model and help constrain the transit parameters.
Including the stellar models in the transit analysis and assuming a circular orbit the solution is much better constrained and we obtain $a/R_*= 4.475^{+0.052}_{-0.17}$, $i= 81.28^{0.36}_{-0.8}$\degree\ and $R_p/R_*=  0.01813^{0.00033}_{-0.00085}$. This solution agrees within $1\sigma$ with the final solution presented by \citet{Leger2009}.

\section{Comparison of LRa01 and LRa06}
\label{comparison}

\begin{figure}
  \centering
  \includegraphics[width=\columnwidth]{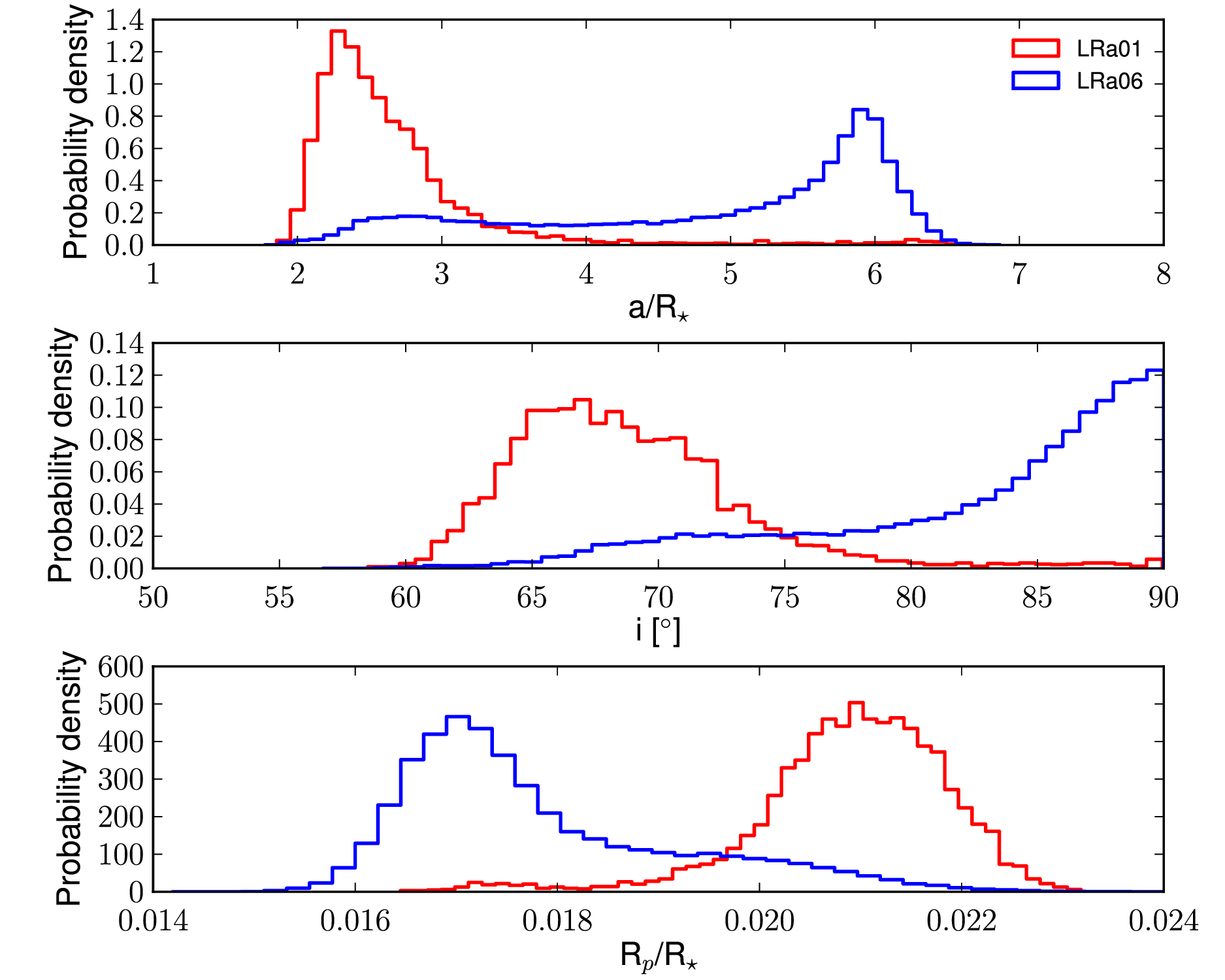}
  \caption{Posterior distribution of the geometric parameters for the separate fits of the LRa01 (red) and the LRa06 (blue) runs. }
  \label{compare}
\end{figure}

\begin{figure}
  \centering
%\vspace{-1cm}
%\hspace{-1cm}
  \includegraphics[width=1.2\columnwidth]{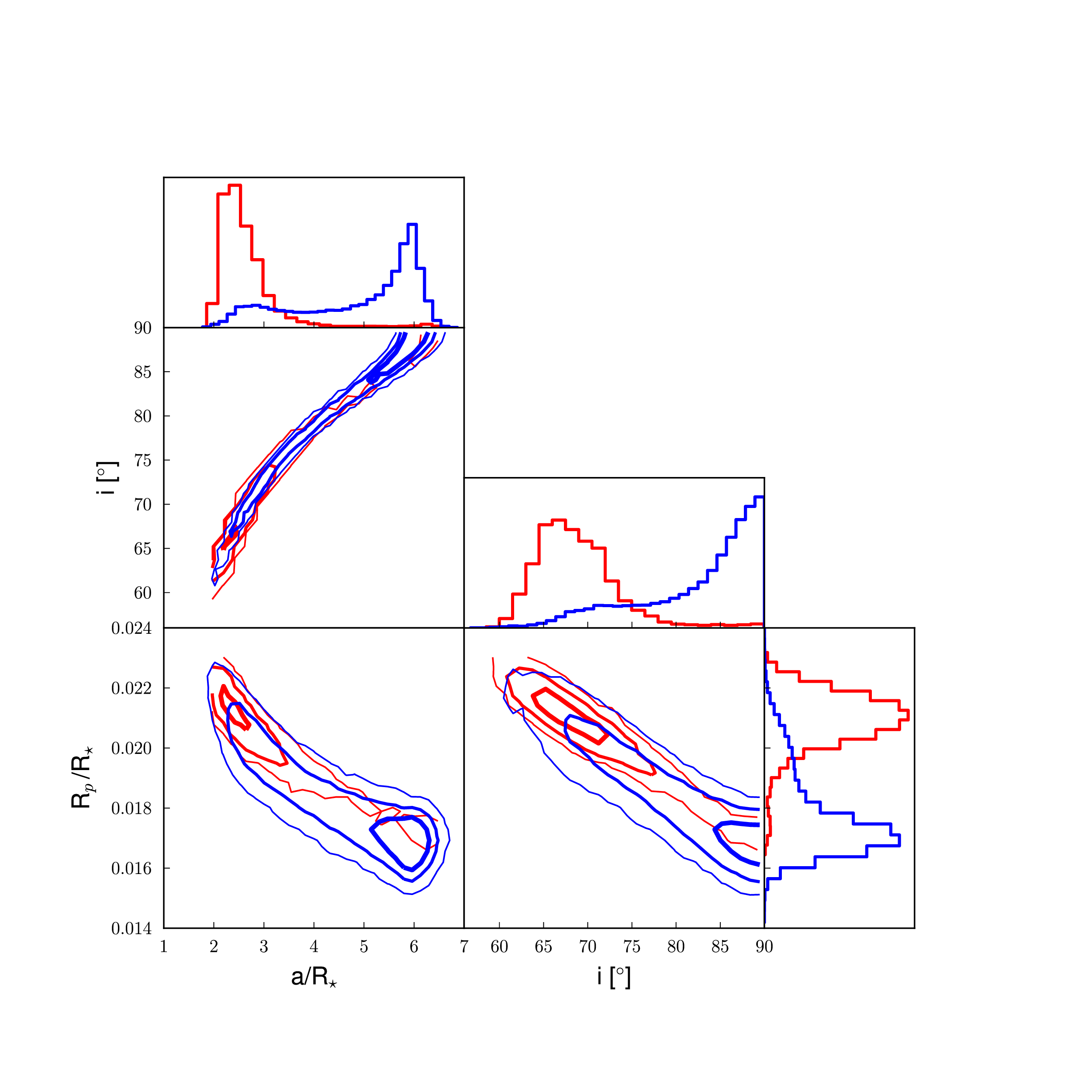}
  \caption{Correlation plots for the  $a/R_*$, $R_p/R_*$ and $i$ derived by the pure geometric fits of LRa01 (red) and LRa06 (blue). For each we show the $1\sigma$, $2\sigma$ and $3\sigma$ contours. }
  \label{correlation}
\end{figure}

As mentioned above when stellar priors are not included, the transit-derived stellar density for LRa06 is much higher than for LRa01. To investigate this inconsistency we compared the posterior distributions of the parameter fits for each  observation. Therefore, we perform a pure geometric fit to the observations of LRa01 following the same procedure as described above for LRa06. The most probable solution (the mode of the distribution) for LRa01 is given in Table~\ref{comptable} together with the results for LRa06 (LRa01 geo and LRa06 geo). The transit-derived stellar density for LRa01 is $0.27^{+0.17}_{-0.10} \rho_{\odot}$, significantly different than for LRa06. The parameter posterior distributions of $a/R_*$, $i$, and  $R_p/R_*$ for LRa01 and LRa06 are shown in Figure~\ref{compare}. It is obvious that the distributions are very different for the two runs. A Kolmogorov-Smirnov test gives a practically null probability that the distributions are drawn from the same sample. However, the difference between the two distributions contains zero at 90\% confidence limit which is consistent with the modes of the two distributions agreeing within 2 sigma. Nevertheless, this difference is worth investigating. A different contamination value related to the sizes of the masks may explain the difference in the depth of the transits. However, instrumental noise is not expected to produce large differences in $a/R_*$ or $i$. To better illustrate the differences between the parameter posterior distributions of LRa01 and LRa06 the correlation plots for $a/R_*$, $R_p/R_*$ and $i$ derived by the pure geometric fits of each run are shown in Figure~\ref{correlation}. This can be directly compared with Figure 18 of \citet{Leger2009} for LRa01.

In Table~\ref{comptable} we also give both parameter solutions found by \citet{Leger2009}. 
 The unconstrained analysis (LRa01 geo \citet{Leger2009} ) agrees well with our geometric solution for LRa01. The results obtained by including the stellar models in the PASTIS fit and assuming a circular orbit for each of the runs are also given in Table~\ref{comptable} (LRa01 star and LRa06 star). The stellar models constrain $a/R_*$ that constrains $R_p/R_*$ and $i$ (Figures \ref{correlation} and \ref{density}). We conclude these agree with each other and agree with the solution of \citet{Leger2009}. This shows that our geometric and full model fit including stellar models are consistent with the method and results of \citet{Leger2009}. The inconsistency is between the geometric solution for LRa01 and the final solution constrained by the stellar models that agree only at $3\sigma$. Hence,  the posterior probability distribution of LRa01 seems to be somehow distorted. It is important to understand this inconsistency to avoid biasing the estimation of the derived systems parameters.

\begin{table}
\centering
\caption{Geometric fitted parameters for LRa01 and LRa06. We also show the results presented by \citep{Leger2009} ([1]) for a simple fit and a fit with a prior on the stellar radius. \label{comptable}}
\begin{tabular}{l c c c}
\hline
Reference &  $a/R_*$ & $R_p/R_*$ & $i$ [\degree]\\
\hline

LRa01 geo [1] & $1.9\pm 0.1$ & $\sim 0.0215$ & $\sim 65$ \\
LRa01 star [1]  & $4.27 \pm 0.2$ & $ 0.0187 \pm 0.003$ & $80.1 \pm 0.3$ \\
LRa01 geo & $2.28^{+0.65}_{-0.12}$  &  $0.02096 \pm 0.0008$  & $67.0^{+5.1}_{-2.7}$\\
LRa01 star & $4.487^{+0.041}_{-0.19}$  &  $ 0.018673^{0.00025}_{-0.00042}$  & $80.82^{0.28}_{-0.74}$\\
LRa06 geo & $5.90^{+0.05}_{-2.8}$ & $0.01702^{0.0023}_{-0.00052}$& $ 90_{-10}$ \\
LRa06 star & $4.475^{+0.052}_{-0.17}$ & $ 0.01813^{0.00033}_{-0.00085}$& $ 81.28^{0.36}_{-0.8}$ \\
\end{tabular}
\end{table}

The cause of the density inconsistency was investigated by \citet{Leger2009} and  two possible causes were suggested. The first one was transit time variations (TTVs) caused by the presence of other planets in the system. This hypothesis was supported by the fact that the derived stellar density was higher if TTVs were accounted for. However, the TTVs timescales were inconsistent with gravitational interaction with other planets and a second hypothesis suggested, that TTVs were induced by stellar activity. \citet{Leger2009} favoured the stellar activity hypothesis as the cause for the degradation of the transit shape resulting on the ingress and egress being less steep than expected. Therefore, the authors chose to fix the stellar radius in the parameter fit process to obtain the system parameters (LRa01 star \citet{Leger2009}) for their adopted solution.

Fortunately, the stellar activity was lower during LRa06 observations. During LRa06, the peak-to-peak variability amplitude of the light curve is half of the value during LRa01. The activity induced RV variations have a maximum amplitude of 53 \ms for the follow-up run of LRa01 while for the simultaneous observations with LRa06 it is 30 \ms \citep{Queloz2009,Haywood2014}. The spectroscopic activity index ($\log R'_{\mathrm{HK}}$) measured in both RV data sets also decreased from $-4.60\pm0.03$ to $-4.73\pm 0.03$. Therefore, it is possible that the stellar activity deformation of the transit shape is higher in LRa01 than in LRa06 which would explain our results. Another hypothesis is that the cause of the shape deformation is instrumental. These two possibilities will be investigated in the next Section~\ref{tests}. However, we start by testing the TTV hypothesis.

\subsection{Transit time variations}

\label{ttvs}

\begin{figure*}
  \centering
  \includegraphics[width=2.\columnwidth]{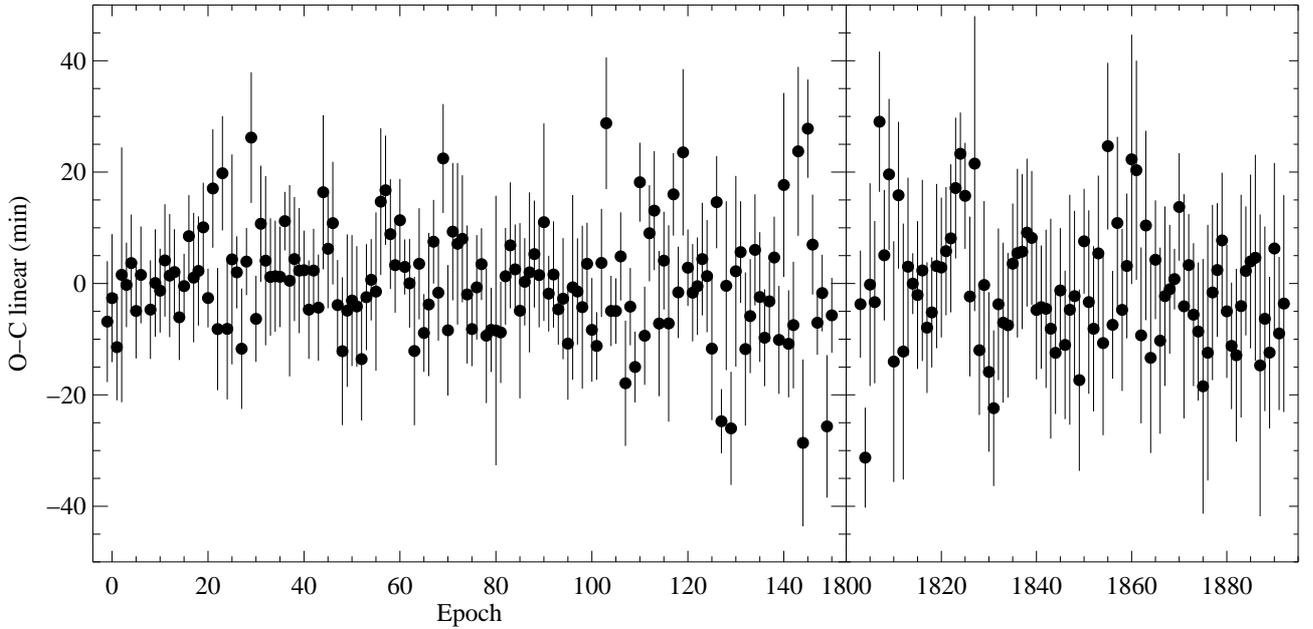}
\caption{Derived TTVs for CoRoT-7b assuming a linear ephemeris for LRa01 (left panel) and LRa06 (right panel).}
\label{ttv}
\end{figure*}

As mentioned before, \citet{Leger2009} reported that the stellar density discrepancy was minimised if the TTVs were accounted for. Therefore, to gain insight into the causes of the density discrepancy we computed the TTVs of both LRa01 and LRa06. Due to the low signal-to-noise ratio of each individual transit of CoRoT-7b, it is challenging to obtain individual transit times. In order to simplify the problem, the transit shape was fixed to the circular final solution presented in Table~\ref{mcmc}.
To obtain the transit times we used a procedure similar to \citet{Barros2011b} but each transit was fitted individually with the fixed shape. We included a linear baseline function in the fit and hence three parameters were fitted for each transit: T0, $F_{out}$ and $F_{grad}$. Due to the poor quality of the light curve we imposed a uniform prior in the transit times, restricting them to be within half of the transit duration from the linear ephemeris. Using a prior with double or half of this size did not significantly alter the results.
 We estimated the transit times with the median of the distribution and the $1\sigma$ limits as the value at which the normalised integral of the distribution equals 0.341 at each side of the median.
The posterior distributions of the transit times are clearly not Gaussian and some present several peaks hence, we caution the reader to an over-interpretation of the individual results and instead prefer to analyse them in a qualitative way.
The residual transit times after removing a linear ephemeris, also called transit timing variations (TTVs) are shown in Figure~\ref{ttv}.

\begin{figure}
  \centering
  \includegraphics[width=\columnwidth]{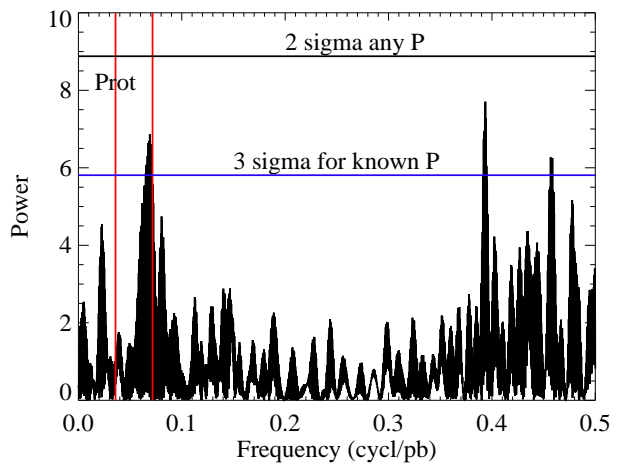}
\caption{Lomb scargle periodogram of the derived TTVs of CoRoT-7b. The rotation period of the star and half of the rotation period of the star are shown as vertical red lines. We over lay the $2~\sigma$ false alarm probability level for any period  (black horizontal line)  and the $3~\sigma$ false alarm probability level for a known period  (blue horizontal line).}
\label{lombscargle}
\end{figure}

We find that the transit times of CoRoT-7b agree well with a linear ephemeris, with a reduced chi-square of 0.95 corresponding to p-value of 29. We also conclude that there is neither a significant period variation nor a transit time variation between the two epochs of the CoRoT observations. To explore periodicities in the TTVs, the Lomb-Scargle periodogram was computed (Figure~\ref{lombscargle}). There is no significant peak at the $2\sigma$ confidence limit (dashed horizontal lines) if we perform a random periodicity search. The conclusion is that we do not detect any transit time variations in CoRoT-7b due to the gravitational influence of other planets.

As mentioned before, two additional planet signatures were found in the radial velocities of CoRoT-7 \citep{Queloz2009, Hatzes2011}. Because both are not in resonance with CoRoT-7b and have relatively low mass, the expected TTVs are very small \citep{Holman2005, Agol2005}. Taking the planetary parameters of \citet{Hatzes2011} and performing Mercury6 simulations \citep{Chambers1999} we find that the expected TTVs due to CoRoT-7c and CoRoT-7d during the observations runs are less than 4 seconds which is much smaller than our timing accuracy if the planets are coplanar. We can exclude a difference in the inclination between the orbits of planet b and planet c higher than 10\degree because this would imply a measurable difference in the transit duration of the planet b \citep{Dvorak2010} which is not seen. From Table~\ref{comptable} we derive a transit duration  of $1.056 \pm 0.017\,$hours for LRa01 and $1.105 \pm 0.035\,$hours for LRa06 when we include stellar models. These are compatible within $1~\sigma$.

 However, if the TTVs are induced by stellar activity they can have periodicities related to the stellar rotation period. For Kepler-17b, that shows clear planet-spot occultation events, the TTVs have a periodicity at half of the rotation period of the star \citep{Desert2011}. In WASP-10b the TTVs periodicity is at the rotation period of the star \citep{Barros2013} and for CoRoT-8b the TTVs periodicity is also related to the rotation period of the star \citep{Borde2010}. In the Lomb-Scargle periodogram of the CoRoT-7b TTVs there is a peak close to half of the rotation period of the star. Although this period is not significant in a blind periodicity search, if we search specifically at the rotation period of the star or half of the rotation period of the star we find that the latter is significant at $3\,\sigma$. Therefore, there is a hint that the transit shapes and times of CoRoT-7b are affected by stellar activity. In the case of planet-spot occultations, the deformation of the transit shape is accompanied by apparent transit timing variations \citep{Alonso2009, Barros2013}.

\section{Distortion of the posteriors of LRa01}

\label{tests}
\subsection{White noise test}

As shown in Section~\ref{whitenoise} the derived parameters from LRa06 are poorly constrained due to the low signal-to-noise of the transits. However the solution is more precise for LRa01 due to the distortion of the transit shape that results in a high impact parameter. The best geometric solution derived from LRa01 is only consistent with the final solution presented in \citet{Leger2009} at $3\,\sigma$. In order to test if the difference in the  posterior distribution shape of the LRa01 relative to LRa06 is caused by the different sampling due to the gaps of cause by the SAA passage we also tested the white noise hypothesis for LRa01.
At the LRa01 observed times we simulated transits with the same transit model shape as used for the LRa06 white noise test (this corresponds to the final solution given in Table~\ref{mcmc}). To the simulated transits we added  white noise with standard deviation corresponding to this run $\sigma=0.00097$. The simulated transits were fitted with the pure geometric PASTIS model. Stellar models were not included in these intermediate tests because they strongly restrict the parameter space, they will only be included in the final analysis. We found that the parameter posterior distributions the simulated transits is the same as for LRa06 implying that the difference in the posterior distribution shape is neither due to the low signal-to-noise of the data nor to the different time of the sampling. The difference between the posterior distributions of $a/R_*$ for the LRa01 and the LRa01 white noise simulations is higher than zero at 95\% confidence limit. Therefore, we conclude that the LRa01 posterior parameter distributions cannot be explained by the assumed transit shape and white noise.

\subsection{Instrumental noise test}

To investigate if the distortion of the posterior distribution of LRa01 was due to instrumental noise, the transit shape mentioned above was also injected in the light curve of a neighbour star in the LRa01 run. This star (CoRoT ID=102727008) was chosen due to the combination of its brightness (V=12.3) and its proximity to CoRoT-7 ($9.5\,$'). The transits were injected in the light curve of 102727008 at the transit times of CoRoT-7b according to a linear ephemeris as before. The posterior distribution of the parameters is similar to the one obtained for the white noise simulations. Hence, we exclude global instrumental noise as the cause of the distortions of the posterior parameter distributions.

\subsection{Red-noise / Out-of-transit variability test}

\begin{figure}
  \centering
  \includegraphics[width=\columnwidth]{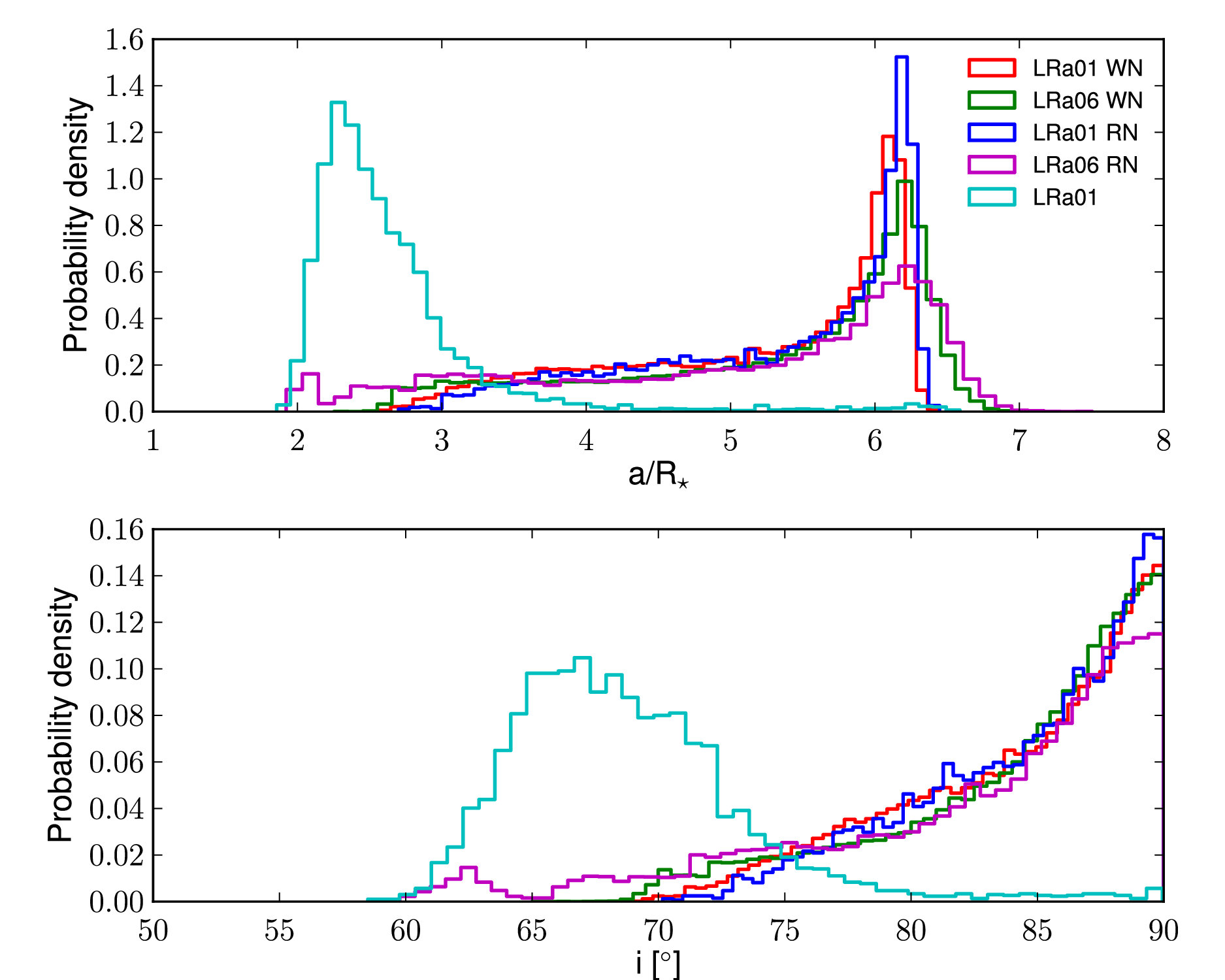}
  \caption{Posterior distribution of $a/R_*$ (top) and $i$ (bottom) for pure geometric fits of the simulated transit model plus white noise (WN) at phase 0, and the transit model plus real data at phase 0.6 (RN) for both LRa01 and LRa06. For comparison we also show the results for the observations of LRa01.}
  \label{rednoise}
\end{figure}

In this section, we investigate if the posterior distribution shape of LRa01 may be caused by non-white noise in the light curve of CoRoT-7. 
This may be due to out-of-transit stellar variability or localised instrumental effects.
A possible way to test this is to inject transits in the real light curve of CoRoT-7 at phases different from the transit phase. Therefore, transits were injected in both CoRoT runs at phases 0.2, 0.4, 0.6 and 0.8 with the same transit model explained above. The simulated transits were fitted with the pure geometric model as explained above. We find that the posterior distributions for the different phases agree well. 
In Figure~\ref{rednoise}, we present a comparison of the posterior distributions of the most problematic parameters, $a/R_*$ and $i$, for transits injected in the two observed light curves of CoRoT-7 at phase 0.6, as an example. In the same figure we also show the analysis of transits injected at phase zero which include only white noise (WN) and the results for the real transits of LRa01 for comparison.

It is evident that the shape of the parameter posterior distributions for all the injected transits light curves is similar but different from the LRa01 light curve. For phases 0.2, 0.6 and 0.8 the difference the posterior of $a/R_*$ from the observed LRa01 is non zero at 95\% confidence but for phase 0.4 the difference is only non-zero at 90\% confidence. This implies that neither the out-of-transit stellar variability nor instrumental red-noise is unlikely to explain LRa01 observations. However, LRa01 data could still be affected by in-transit stellar activity, more specifically planet-spot occultation events. Due to the low signal-to-noise ratio of each individual transit, fitting a spot model is not feasible, so this hypothesis cannot be tested directly.

\subsection{Test out-of-transit flux}
Our previous tests suggest that there is an extra noise component in the LRa01 not present at other transit phases. This degrades the transit shape which appears more V-shaped.
To test if this is related to the visible spots we divided transits according to their out-of-transit flux level and use the median as a threshold separately for LRa01 and LRa06.
Each of the selections were fitted with the pure geometric models as before.

\begin{figure}
  \centering
  \includegraphics[width=\columnwidth]{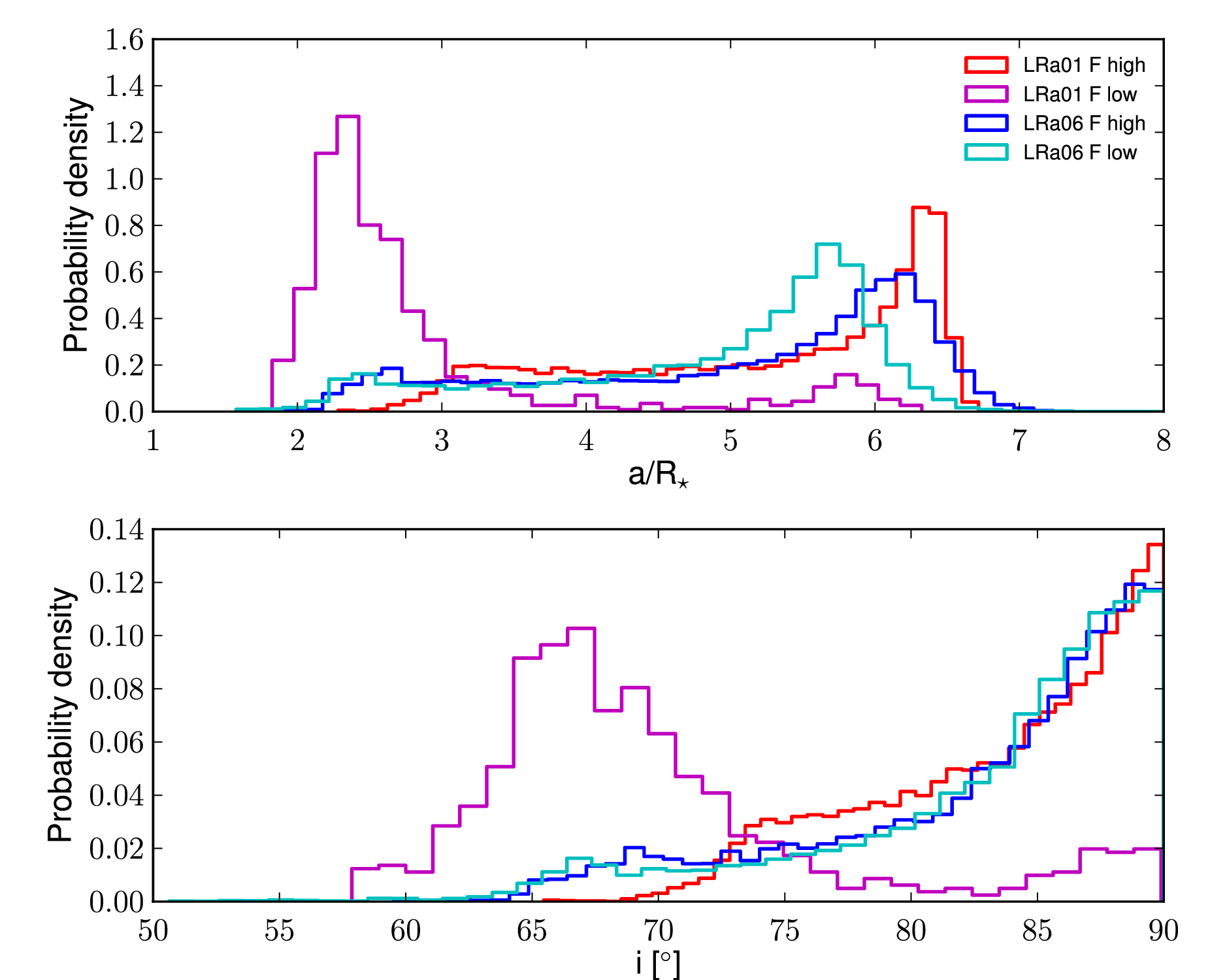}
  \caption{Posterior distribution of $a/R_*$ (top) and $i$ (bottom) for pure geometric fits of the transit selections based on the out of transit flux. The transits at high flux levels are shown in red for LRa01 and blue for LRa06 while the one at the low flux levels are shown in magenta for LRa01 and cyan for LRa06.}
  \label{dist_fout}
\end{figure}

In Figure~\ref{dist_fout}, we show a comparison of the posterior distributions of $a/R_*$ and $i$ for the high and low flux levels of each run. We find that the posterior distributions are all similar except for the selection of the low flux level of LRa01. We conclude that the low flux level transits of LRa01 appear to have more extra noise than the other samples. As mentioned above, in CoRoT-7 the activity is dominated by spots. Low flux levels imply a larger spot coverage and hence a higher probability of spot occultation events during transit. In LRa06 the amplitude of the flux variability was half of the one found in LRa01. If we assume that the maximum flux in both light curves (LRa01 and LRa06) is the same, LRa06 has similar spot coverage than the high flux level selection of LRa01 which could explain our results.

 \subsection{Selecting transits}
 
The effect of stellar activity on individual transits is not clear because of the low signal-to-noise ratio. Our tests, however, show that the deformation of the transits is not related to out-of-transit stellar variability or instrumental noise. Both the periodicity found in the TTVs related to half of the rotation period of the star and the distortion of the posterior probability distributions for lower flux levels support that the cause of the extra noise is spot occultation events.

To avoid biasing the final derived parameters we attempted to select the transits that would be less affected by this noise component. We tried selections based on the chi-squared of individual fits and the measured transit times. However, these selections assume a specific transit shape and can bias the depth of the transit. Therefore, we opt to select transit on the basis of out-of-transit flux that does not assume a transit shape. For the final result we combined the high flux level selection of LRa01 with the LRa06 full dataset.These two light curves show almost the same amplitude of flux variability and similar parameter posterior distributions.

\section{Results}

\label{results}

In our final analysis of the data we combined all transits of LRa06 with transits of LRa01 that have out-of-transit flux level higher than the median flux. These were fitted with PASTIS together with the RVs and with the stellar models as described in Section~\ref{modelfull} to derive full system parameters. From Table~\ref{comptable} we conclude that the derived $R_p/R_*$ for both runs agree within $1\sigma$ when stellar models are included. Therefore, there is no significant difference in contamination between both runs. The mode of the posterior distribution and the 68.3\% central confidence interval of the derived system parameters are presented in Table~\ref{mcmc}. Since the value of the eccentricity is consistent with zero, the circular orbit  solution is also presented in the same Table. The determined stellar parameters were combined with the CoRoT-7 photometric magnitudes from APASS \footnote{http://www.aavso.org/apass}, 2MASS \citep{Skrutskie2006}, and WISE catalogues \citep{Wright2010} to derive the distance to the system which is also given in Table~\ref{mcmc}.

\begin{table*}
\centering
\caption{CoRoT-7 system parameters.}
\begin{minipage}[t]{\textwidth}
\setlength{\tabcolsep}{3.0mm}
\renewcommand{\footnoterule}{}
\begin{tabular}{l c c}
\hline\hline
                                                 &  eccentric               & circular \smallskip\\
& & adopted \smallskip \\

Planet orbital period, $P$ [days]$^{\bullet}$      & 0.85359163 $\pm$ 5.8e-7          & 0.85359159 $\pm$ 5.7e-7 \\
Mid-transit time, $T_{c}$ [BJD]$^{\bullet}$         & 2454398.07741 $\pm$ 6.9e-4     & 2454398.07756$^{+4.5e-4}_{-7.4e-4}$ \\
$cov(P,T_{c})$ [days$^2$]                         & -2.37$\times 10^{-10}$                 & -2.02$\times 10^{-10}$      \\
Orbital eccentricity, $e$$^{\bullet}$              & 0.137$^{+0.094}_{-0.053}$ $\textless$ 0.32$^{\dagger}$&                           \\
Argument of periastron, $\omega$ [deg]$^{\bullet}$ & 81 $\pm$ 30                            &                           \\
Orbit inclination, $i$ [deg]$^{\bullet}$           & 81.20$^{+1.5}_{-0.44}$                    & 80.78$^{+0.51}_{-0.23}$      \\
Orbital semi-major axis, $a$ [AU]                & 0.017027$^{+1.6e-4}_{-4.7e-5}$             & 0.017016$^{+1.7e-4}_{-3.6e-5}$ \\
semi-major axis / radius of the star, $a/R_{\star}$& 4.469$^{+0.087}_{-0.040}$         & 4.484 $\pm$ 0.070 \smallskip \\

Radius ratio, $k=R_{p}/R_{\star}$$^{\bullet}$         & 0.01721 $\pm$ 0.00060   & 0.01784 $\pm$ 0.00047   \\
Linear limb darkening coefficient, $u_a$          & 0.517 $\pm$ 0.015   & 0.515 $\pm$ 0.014        \\
Quadratic limb darkening coefficient, $u_b$       & 0.187 $\pm$ 0.010 & 0.188 $\pm$ 0.011     \\
Transit duration, $T_{14}$ [h]                     & 1.049$^{+0.035}_{-0.019}$  & 1.059 $\pm$ 0.026   \\
Impact parameter, $b$                             & 0.594$^{+0.066}_{-0.14}$  & 0.713$^{+0.017}_{-0.026}$ \smallskip \\

Radial velocity semi-amplitude, $K$ [\ms]$^{\bullet}$   & 3.85 $\pm$ 0.60   & 3.94 $\pm$ 0.57 \smallskip\\

Effective temperature, $T_{\mathrm{eff}}$[K]$^{\bullet}$   & 5267 $\pm$ 60         & 5259 $\pm$ 58             \\
Metallicity, $[\rm{Fe/H}]$ [dex]$^{\bullet}$            & 0.137 $\pm$ 0.059 & 0.138 $\pm$ 0.061         \\
Stellar Density, $\rho_{\star}$ [$\rho_\odot$]$^{\bullet}$ & 1.665 $\pm$ 0.080 & 1.671 $\pm$ 0.073      \\
Star mass, $M_\star$ [\Msun]                           & 0.915 $\pm$ 0.019      & 0.913 $\pm$ 0.017 \\
Star radius, $R_\star$ [\Rsun]                         & 0.818 $\pm$ 0.016      & 0.820 $\pm$ 0.019 \\
Deduced stellar surface gravity, $\log$\,$g$ [cgs]    & 4.572 $\pm$ 0.011     & 4.573 $\pm$ 0.011  \\
Age of the star [$Gyr$]                               & 1.32 $\pm$ 0.76       & 1.32 $\pm$ 0.75      \\

Planet mass, $M_{p}$ [M$_{\oplus}$ ]                    & 5.55 $\pm$ 0.85      & 5.74 $\pm$ 0.86  \\
Planet radius, $R_{p}$[R$_{\oplus}$]                    & 1.528 $\pm$ 0.065    & 1.585 $\pm$ 0.064 \\
Planet mean density, $\rho_{p}$ [$g\;cm^{-3}$]         & 8.1 $\pm$ 1.6        & 7.5 $\pm$ 1.4         \\
Planet surface gravity, $\log$\,$g_{p}$ [cgs]         & 3.348 $\pm$ 0.068    & 3.332 $\pm$ 0.065 \\
Planet equilibrium temperature$^\ast$, $T_{eq}$ [K]    & 1759 $\pm$ 30        & 1756 $\pm$ 27       \smallskip\\

\hline
Distance of the system [pc]$^{\bullet}$                &  154.2 $\pm$ 4.7     & 153.7 $\pm$ 4.5   \smallskip \\

\hline
Adopted Planet mass, $M_{p}$ [M$_{\oplus}$ ]   & \multicolumn{2}{c}{4.73 $\pm$ 0.95 \citep{Haywood2014}  } \\

Adopted Planet mean density, $\rho_{p}$ [$g\;cm^{-3}$]         &   \multicolumn{2}{c}{6.59 $\pm$ 1.5 \citep{Haywood2014}}\\
\hline
\hline
\vspace{-0.5cm}
\end{tabular}
\begin{list}{}{}
\item $^{\bullet}$ MCMC jump parameter. $^{\dagger}$ upper limit, 99\% confidence. $^{\ast}$ $T_{eq}=T_{\mathrm{eff}}\left(1-A\right)^{1/4}\sqrt{\frac{R_\star}{2 a}}$, using an albedo $A=0$.
\end{list}
\end{minipage}
\label{mcmc}
\end{table*}

\begin{figure*}
\centering
\includegraphics[width=0.45\textwidth]{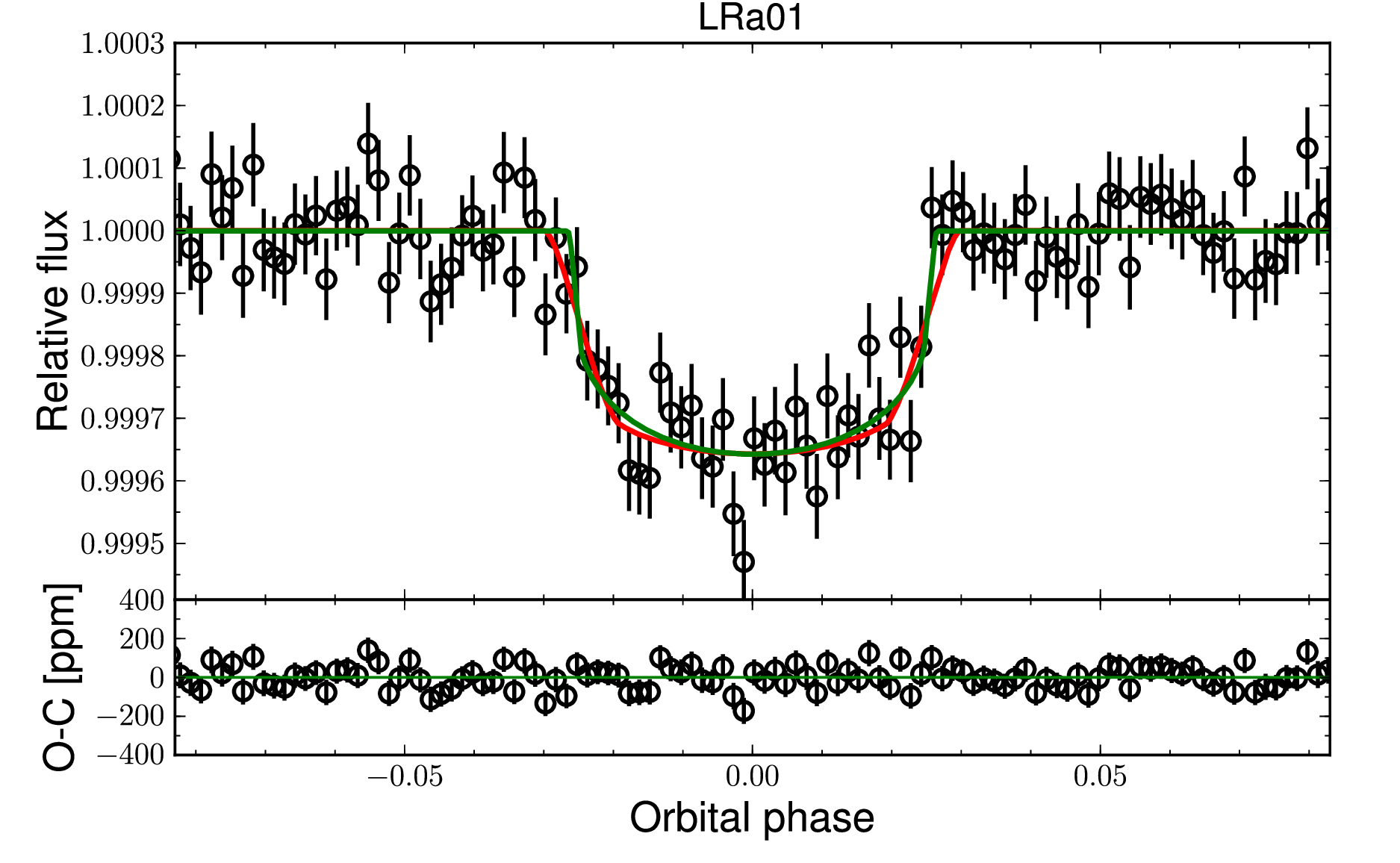}
\includegraphics[width=0.45\textwidth]{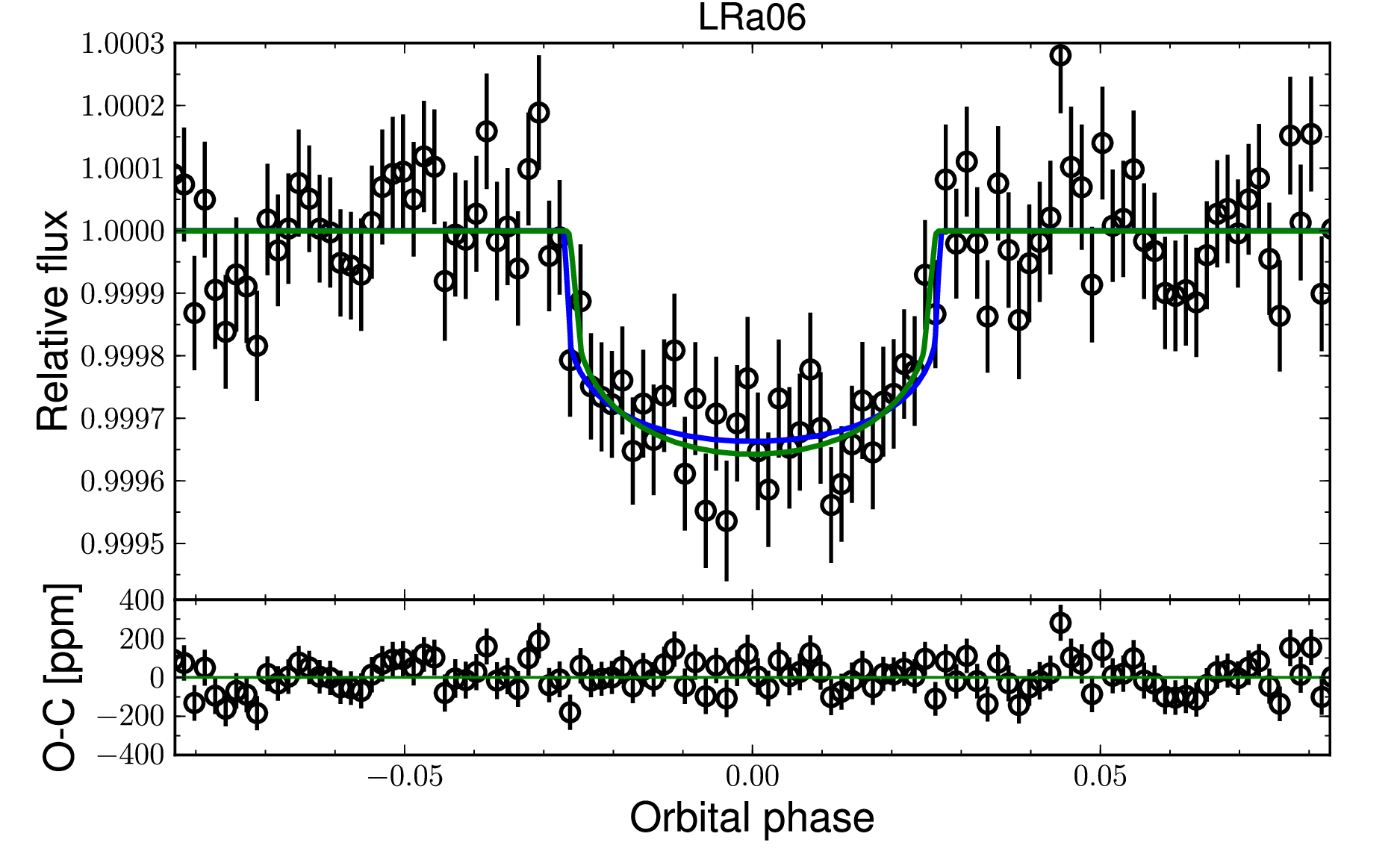}
\caption{Phase folded light curve of CoRoT-7b binned for clarity. In green we show the best transit model derived from fitting simultaneously the CoRoT light curves and HARPS RVs that corresponds to the circular solution given in Table~\ref{mcmc}. We also show the best pure geometric transit model for LRa01 (red) and LRa06 (blue) corresponding to the solutions presented in Figure~\ref{compare} and Table~\ref{comptable}.}
\label{lctransit}
\end{figure*}

\begin{figure}
  \centering
  \includegraphics[width=\columnwidth]{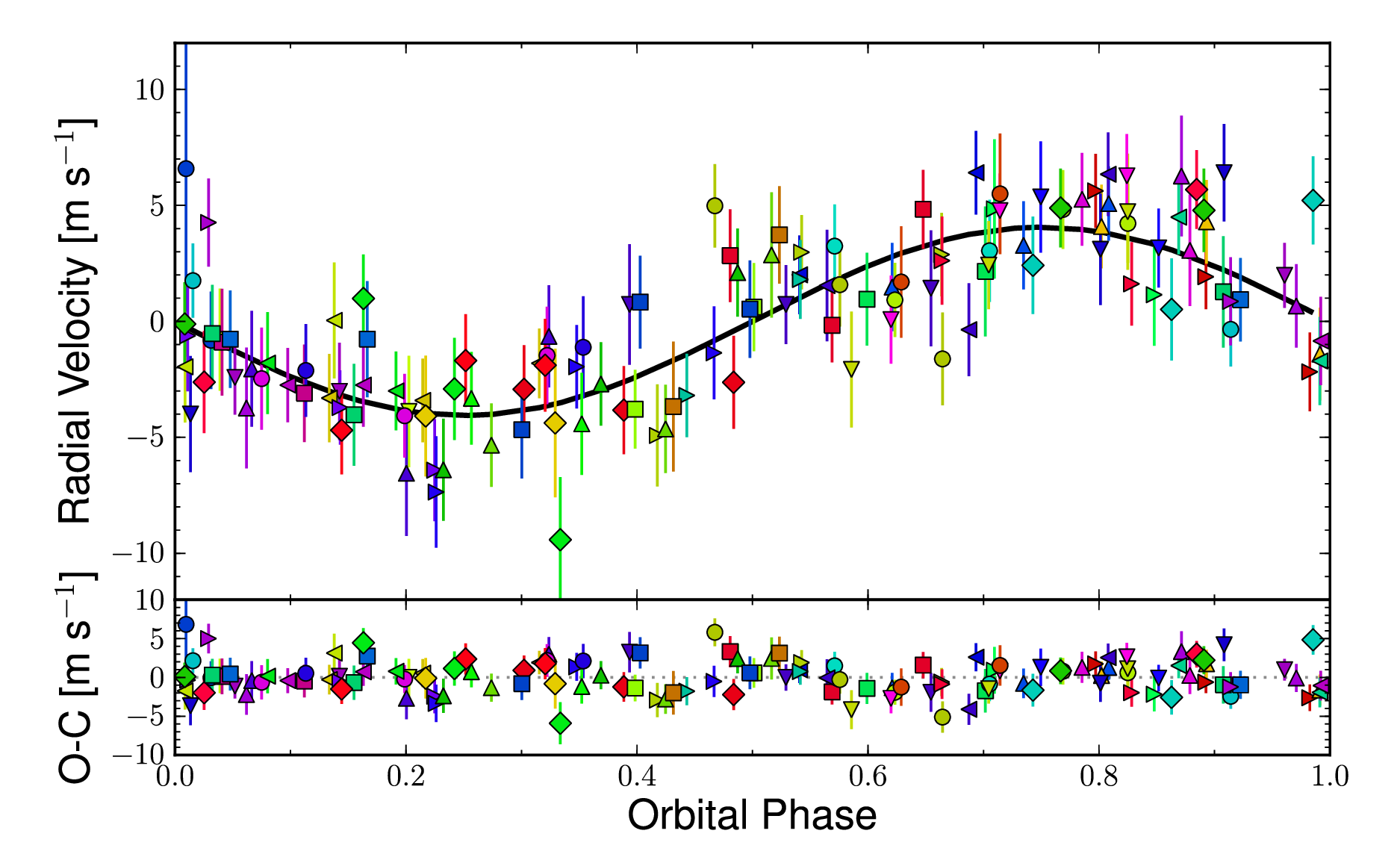}
  \caption{Phase folded radial velocities. The different nights were plotted with a different combination of colour and symbol}
  \label{rvplot}
\end{figure}

We performed a Bayesian model comparison between the circular and the eccentric orbit model. In order to estimate the evidence, we use the \citet{Chib2001} method, obtaining  $\log_{10}$ of the Bayes factor of the circular over the eccentric orbit model of $3.0 \pm 2.4$. We conclude that there is no preference over both models \citep{Diaz2014}. For the sake of simplicity we adopt the circular orbit solution.  We note that the results of both solutions are within $1 \sigma$.
A full discussion on the eccentricity will be presented elsewhere  (\citealt{Haywood2014}, Hatzes, A. et al. in prep).
We show the adopted circular model overlayed on the phase folded transits of CoRoT-7b, in Figure~\ref{lctransit}. In Figure~\ref{rvplot} we show the phase folded radial velocities.

Our derived stellar parameters,  $ \rho_*= 1.671 \pm 0.073  \rho_\odot\,$, $R_*= 0.820 \pm 0.019 \,$ \Rsun\ and  $M_*=  0.913 \pm 0.017\,$ \Msun\  agree well with those of \citet{Bruntt2010b},  $ \rho_*= 1.65 \pm 0.15\,  \rho_{\odot}$, $R_*=0.82 \pm 0.04\,$ \Rsun\ and  $M_*=0.91 \pm 0.03 \,$\Msun.

The measured planet-to-star radius, $ R_p/R_* =  0.01784 \pm 0.00047 \,$ and planetary radius, $R_p= 1.585 \pm 0.064\,$ \rearth\ agree well with the previous estimate \citep{Leger2009}. 
 The derived planet mass, $M_p=  5.74 \pm 0.86\,$ \mearth\, agrees with the previous estimates by \citet{Queloz2009,Hatzes2010, Hatzes2011,Boisse2011, Haywood2014}. This implies a planetary density of $ 1.35 \pm 0.25\,  \rho_{\oplus}$ or $ 7.5 \pm 1.4\, g/cm^3$.

 For high signal-to-noise light curves where the limb darkening can be fitted a discrepancy between the fitted values and the tabulated values from stellar atmospheric models was reported \citep[e.g.][]{Claret2009,Barros2012}. This can be a source of systematic errors in the derived planetary parameters \citep{Csizmadia2013}. The fitting of the limb darkening parameters, however is not feasible in our case. We tried fitting a linear limb darkening law and found  $ \gamma_1 \sim 1$ but it is not well constrained. The derived planetary radius for this fit was smaller by $0.35 \sigma$. Therefore, it can be concluded that fixing the limb darkening does not significantly affect our results.

 The stellar age derived by gyrochronology and by activity-age relations (Age$=[1.2,2.3]\,$Gyr, \citealt{Leger2009, Bruntt2010b})  was used to constrain the stellar models in previous studies. Also in our analysis we constrained the stellar age to be less than $3\,$Gyr. However, if stellar age is not constrained, we find the most probable stellar age to be $8.4^{+5.0}_{-3.3}\,$Gyr and the derived stellar properties are $ \rho_*=1.34 \pm 0.13\, \rho_{\odot}\,$, $R_*= 0.873 \pm 0.034\,$ \Rsun\ and  $M_*=0.865 \pm 0.050\,$\Msun. Since the stellar parameters constrain the transit parameters this has high impact into the derived planetary parameters $R_p=1.94 \pm 0.10\,$ \rearth, $M_p= 5.62 \pm 0.75$  \mearth\ and   $\rho_{p}=  3.81  \pm 0.78\, g/cm^3$. A better constrain on the stellar density for example with asteroseismology would significantly improve the accuracy of the planetary parameters.

\subsection{Planetary density}
\begin{figure}
  \centering
  \includegraphics[width=\columnwidth]{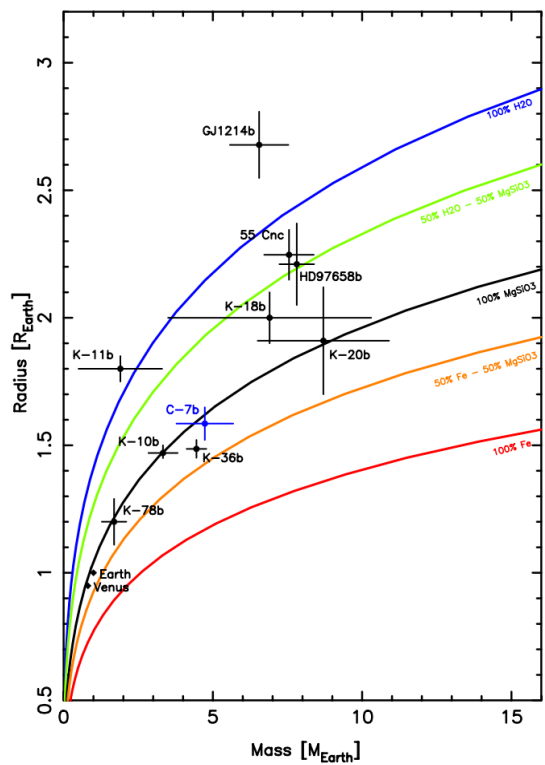}
  \caption{Mass-radius diagram for low mass planets showing the position of CoRoT-7b with $M_p= 4.73 \pm 0.95\,$ \mearth  \citep{Haywood2014} and the derived the radius for the selected sample, $R_p=  1.585 \pm 0.064\,$  \rearth\   (blue). We show the position of Earth and Venus (diamonds) for comparison. The solid lines show the mass and radius for planets with different compositions according to the models of \citet{Zeng2013}.}
  \label{pdens}
\end{figure}

As mentioned before, the uncertainty on the mass of CoRoT-7b prevented constraining its bulk composition in previous works. The lower mass estimate from \citet{Pont2011} allows a gaseous composition. However, all the other mass estimates imply a rocky composition \citep{Valencia2010, Valencia2011,Barnes2010,Leitzinger2011} between 'Earth-like' (33\% iron and 67\% silicate mantle) and 'Mercury-like' (63\% iron and 37\% silicate mantle). The new radial velocity data set allowed a better constraint on the planetary mass. 
As mentioned above, the derivation of the planetary mass is the subject of companion papers (\citealt{Haywood2014}, Hatzes, A. et al. in prep). The RVs were included in our analysis to confirm that the eccentricity is not significant given that a better accuracy on the eccentricity can be achieved when a combined fit of the transits and RVs is performed  \citep[e.g.][]{Barros2011a}. An accurate eccentricity is needed to derive accurate transit parameters. Using our simple method to filter the activity in the RVs we found no significant eccentricity and the derived planetary mass agrees well with the results by \citet{Haywood2014}. \citet{Haywood2014} use the light curve to correct the radial velocities with the ff’ method of \citet{Aigrain2012} combined with a Gaussian process that has the same covariance properties as the light curve.  We show these two very different techniques result in values of the planetary mass and eccentricity  that are in agreement at $1\sigma$. This contrasts with previous estimates of the mass of CoRoT-7b when the stellar activity was higher. Since  the ff’ method is a phenomenological method we adopted their derived planetary mass  $M_p= 4.73 \pm 0.95\,$ \mearth instead of our value using a simple filtering technique. Combining the new value for radius with the mass derived by \citep{Haywood2014} we obtain a planetary density of $\rho_{p} = 6.59 \pm 1.5\,   g/cm^3$ .
This implies that CoRoT-7b is slightly more dense than
the Earth  $\rho_{p} = 1.19 \pm 0.27\,  \rho_{\oplus}$. Figure~\ref{pdens} shows the position
of CoRoT-7b on a mass-radius diagram alongside
other exoplanets for which mass and radius have
been measured.
According to composition models by
\citet{Zeng2013} for solid planets, CoRoT-7b could be composed of silicates combined with water ice or iron. These are also in agreement with the models of  \citet{Wagner2011}. In this case any water ice is mixed with the silicates and does not imply the planet has an atmosphere. Hence, the planetary parameters are compatible with a rocky composition.

 However, the existence of an atmosphere would introduce a degeneracy in the models. The short orbital period and high equilibrium temperature of the planet,  $T_{eq}= 1756 \pm 27 \,$K might be hard to reconcile will an atmosphere dominated by a volatile gas, H2O. The stability of similar Water ocean-planet was studied by \citet{Selsis2007} that derived a lower limit for the lifetime of atmosphere of planets under the erosion of Extreme UV and stellar wind life of the star. According to figure 4 of \citet{Selsis2007} it is possible that for the derived parameters of CoRoT-7 the atmosphere would have been eroded. However, the contrary cannot be excluded since no upper limit on the lifetime was presented. A similar conclusion was found by \citet{Valencia2011} that shows that in CoRoT-7b the age of the system is similar to the timescale of the evaporation of water vapour. The presence of an atmosphere will be clarified with future observations with JWST \citep{Samuel2014}. A better constrain on the stellar and planetary parameters are needed to obtain further insight into the composition of CoRoT-7b.

\section{Discussion}

 We have showed that the transit-derived stellar density for each of the CoRoT-7b observations is different from the spectroscopic derived density, for LRa01 it is lower while for LRa06 it is higher. The comparison between the transit derived stellar density and the spectroscopic derived value has long been used as a blend test in transit surveys \citep{Cameron2007,Tingley2011}. Furthermore, it has also been used to estimate the orbital eccentricity directly from the transit for some Kepler candidates by \citet{Dawnson2012}, the technique was first suggested by \citet{Ford2008b}. Recently, \citet{Kipping2014} discusses five other effects that would lead to the transit derived stellar density being different from the true stellar density.
 In this paper we show two other effects that can lead to a poor precision or accuracy of the stellar density from the transit, respectively the poor resolution of the egress/ingress time and transit spot occultation events that are not resolved in a single transit.

To solve the poor resolution of the egress/ingress time we have used the spectroscopic derived stellar properties (\logg, [Fe/H] and  \teff) together with stellar models to help constrain the transit shape. This allows to constrain the system parameters but assumes that the other effects are negligible. 

In high signal-to-noise light curves, transit spot occultation events are obvious, they can be used to derive spot properties and/or the angle between the stellar rotation and the planet orbit \citep{Wolter2009,Silva-Valio2010,Sanchis-Ojeda2011}. However, in some cases it is non-trivial to identify the effect of spot occultation events in transits, specially during ingress and egress as shown by \citet{Barros2013}. Once the spot occultation events are identified there are several ways to account for them, for example, masking the affected transit phase, simultaneously fitting a occultation spot model or rejecting affected transits. However, when the individual transits are not resolved like for the case of CoRoT-7b presented here, it was not possible to identify affected transits. The only option we found was rejecting transits based on the out-of-transit flux which is related to the spot coverage.

Assuming the distortion of the posterior is significant and related to spot occultation events, these could have affected the radius estimate. For example, for CoRoT-2b it has been shown that spot occultation events lead to an underestimation of the planet-to-star radius  \citep{Silva-Valio2010}. However, in our case case we found no significant difference in the derived planet-to-star radius. If we consider all the transits we obtain a value of the planet-to-star radius consistent within one sigma with value obtained with the selection.

The transit chord covers latitudes between 38 and 41.6 degrees in stellar surface. Sun spots are located at latitudes lower than 30 degrees, however, for stars more active than the sun the spots can reach higher latitudes \citep{Moss2011}.
Besides a difference in the stellar activity level between LRa01 and LRa06 the spot mean latitudes have also changed (\citealt{Lanza2010} and Lanza et al. in prep). This can explain the difference of effect of activity in the transit derived parameters between both runs.

Besides the bias in the parameter estimation due to spot occultation events, the stellar activity in CoRoT-7 can also introduce a bias due to out-of-transit flux variations \citep{Czesla2009}. Assuming the stellar variability in CoRoT-7 is dominated by cool spots \citep{Queloz2009,Lanza2010}, the planet-to-star radius is overestimated.  Using equation 19 of \citet{Kipping2014}, for a stellar variability amplitude of 2\% we estimate the ``true''  planet-to-star radius is $0.99 \times$ the observed value which implies an overestimation of the planetary radius of $0.5\%$. Furthermore, spots that do not produce significant rotational flux modulation will lead to a higher overestimation of the planetary radius that could only be detected by long term monitoring of the star.

\section{Conclusions}
We present new photometric observations of CoRoT-7 with the CoRoT satellite during the LRa06 run. These were combined with the previous CoRoT-7 observations during the LRa01 run and radial velocity observations obtained with HARPS and fitted with the PASTIS code. Due to the lower activity level of CoRoT-7 during LRa06, we were able to disentangle the effects of activity and the difficulty of differentiating the ingress/egress time in shallow transits with low signal-to-noise.

 For pure geometric fits, we find a difference in the transit parameter posterior distribution of the two CoRoT-7 runs. For LRa06 the results are consistent with what would be expected by white noise while for LRa01 we show that the transit parameter posterior distribution is distorted implying a transit shape deformation. 
 
 To investigate the degradation of the transit shape in LRa01, we performed several tests that excluded the low signal-to-noise of the transit and out-of-transit variability due to either activity or instrumental noise as causes of the transit shape deformation. This suggested that the transit shape deformation could be due to planet-spot occultation events and is supported by the fact that the transit times show a periodicity related to half of the rotation period of the star. Moreover, the distortion of the posterior happens only for transits with out-of-transit flux lower than the median level in LRa01. To avoid biasing the estimation of the parameters these transits were discarded in our analysis.

During the observations of LRa06 the star was in a lower activity level and the spot latitudes had probably migrated with respect to LRa01 (\citealt{Lanza2010} and Lanza et al. in prep). Consequently, the transit shape was less affected by activity and the parameter posterior distributions are consistent with the derived transit model for white noise dominated data. However, due to the low signal-to-noise of the transits the ingress/egress time is not well defined leading to a poor constraint on the inclination. Therefore, stellar models were included in our transit fitting procedure in order to constraint the system parameters.

The transits with out-of-transit flux higher than the median level in LRa01 were modelled simultaneously with all LRa06 data,  the HARPS radial velocity data and stellar models to derive system parameters using PASTIS.

We obtained a planetary mass of  $M_p=  5.74 \pm 0.86\,$ \mearth\ supporting the previous values by \citet{Queloz2009,Hatzes2010, Hatzes2011,Boisse2011,Haywood2014}.  The derived planetary radius,  $R_p=   1.585 \pm 0.064 \,$   \rearth\ agrees well with previous results. Combining our planetary radius with the planetary mass derived by \citet{Haywood2014} we obtain a planetary density of  $  1.19 \pm 0.27\,  \rho_{\oplus}$ which implies a composition with a significant amount of silicates.

Our results illustrate the difficult in recognising the effects of activity in small planet transits. However, in this case we find that there is no significant effect on the derived planetary parameters. This could be due to the solution being dominated by the prior on the stellar density using stellar models. Therefore, our results are strongly dependent on the stellar parameters and the age limit assumed and would benefit from tighter constraints on the stellar properties.

\begin{acknowledgements}
We thank the referee Roberto Sanchis-Ojeda for his careful review and constructive comments, which significantly contributed to improving the quality of the paper.
The CoRoT space mission, launched on December 27th 2006, has been developed and is operated by CNES, with the contribution of Austria, Belgium, Brazil , ESA (RSSD and Science Programme), Germany and  Spain.
 We acknowledge the PNP of CNRS/INSU, and the French ANR for their support. The team at LAM acknowledges support by grants  98761 (SCCB) and 251091 (JMA). RFD was supported by CNES via the its postdoctoral fellowship program. AS acknowledges the support of the European Research Council/European Community under the FP7 through the Starting Grant agreement number 239953.
This research was made possible through the use of the AAVSO Photometric All-Sky Survey (APASS), funded by the Robert Martin Ayers Sciences Fund.
This publication makes use of data products from the Two Micron All Sky Survey, which is a joint project of the University of Massachusetts and the Infrared Processing and Analysis Center/California Institute of Technology, funded by the National Aeronautics and Space Administration and the National Science Foundation.
This publication makes use of data products from the Wide-field Infrared Survey Explorer, which is a joint project of the University of California, Los Angeles, and the Jet Propulsion Laboratory/California Institute of Technology, funded by the National Aeronautics and Space Administration.
 The team at IAC acknowledges support by grant AYA2012-39346-C02-02 of the Spanish Ministerio de Economi\'a y Competividad.
 The German CoRoT team (TLS and University of Cologne) acknowledges
DLR grants 50OW0204, 50OW0603, and 50QM1004.
\end{acknowledgements}

\bibliographystyle{aa} 
\bibliography{susana}

\end{document}